\documentclass[amsmath,trackchanges,twocolumn]{aastex702}

\usepackage{amsmath}
\usepackage{graphicx}
\usepackage{placeins}
\usepackage[all]{hypcap}

\begin{document}

  \title{Constraining the FRB Central Engine from Flux Variations of Persistent Radio Sources:\\ FRB 20240114A as the First Attempt}

  \author[orcid=0009-0004-9433-9351]{Hirotaka Matsumiya}
  \affiliation{Department of Astronomy, School of Science, The University of Tokyo, Bunkyo-ku, Tokyo 113-0033, Japan} 
  \email{hirotaka-ma2@g.ecc.u-tokyo.ac.jp}

  \author[orcid=0000-0003-4299-8799]{Kazumi Kashiyama} 
  \affiliation{Astronomical Institute, Graduate School of Science, Tohoku University, Sendai 980-8578, Japan}
  \email{kashiyama@astr.tohoku.ac.jp}

  \author[orcid=0000-0002-6156-6783]{Tomonori Totani}
  \affiliation{Department of Astronomy, School of Science, The University of Tokyo, Bunkyo-ku, Tokyo 113-0033, Japan}
  \affiliation{Research Center for the Early Universe, School of Science, The University of Tokyo, Bunkyo-ku, Tokyo 113-0033, Japan}
  \email{totani@astron.s.u-tokyo.ac.jp}

  \begin{abstract}
    Some repeating fast radio bursts (FRBs) are associated with persistent radio sources (PRSs), likely driven by the long-term activity of a central engine.
    We investigate FRB 20240114A---currently the most active repeating FRB---to explore the link between its burst activity and the flux variability of its associated PRS. 
    We model electrons responsible for FRB emission being continuously injected from a central magnetar into its magnetar wind nebula (MWN), where synchrotron radiation produces the PRS. 
    Comparing our model with the latest PRS spectra and flux variations, we constrain the physical properties of the FRB source. 
    We find that the total energy injected into the MWN during FAST observations from January to August 2024 is $\sim 10^{48} \, \rm {erg}$, consistent with the magnetic energy budget of a magnetar with an internal magnetic field of $\sim 10^{16} \, \rm G$. 
    The bulk Lorentz factor of the FRB-emitting plasma is mildly constrained to $\lesssim 10^3$, and FRB 20240114A is inferred to be powered by a young magnetar aged $\sim 10$--$20 \, \rm {yr}$. 
    These results show that the PRS flux variability effectively probes FRB emission properties. 
    We predict that while PRS luminosity increases rapidly with enhanced FRB activity, it will slowly fade over decades via adiabatic cooling after FRB activity ceases.
    Energetics suggest that the hyper-active phase observed in FRB 20240114A (burst rate $\sim 700 \, \rm hr^{-1}$) is a rare state with a duty cycle of $\lesssim 10^{-2}$, which is broadly consistent with its detection rate in current FRB-PRS samples. 
   \end{abstract}

  \section{Introduction}
  \label{sec:intro}

    Fast radio bursts (FRBs) are bright, millisecond-duration radio transients with flux densities reaching the Jansky level, originating from extragalactic distances (\citealt{lorimer_2007}; \citealt{petroff_2022}; \citealt{zhang_2023a}).
    However, their detailed emission mechanism remains poorly understood.
    To date, more than 3,000 FRB sources have been discovered, exhibiting a wide range of temporal and spectral properties (\citealt{chime_2026}). 
    Among them, a subset shows repeating activity, beginning with the discovery of the first repeater, FRB 20121102A (\citealt{spitler_2016}), and more than 80 such sources are now known.

    The detection of FRB 20200428 from a Galactic magnetar (SGR 1935+2154) (\citealt{bochnenek_2020}; \citealt{chime_2020}) has provided compelling evidence that at least a fraction of FRBs are powered by magnetars.
    Those bursts from SGR 1935+2154 have also been accompanied by X-ray bursts (XRBs), indicating that FRBs can be associated with multi-wavelength emission (\citealt{mereghetti_2020}; \citealt{li_2021}; \citealt{ridnaia_2021}; \citealt{tavani_2021}; \citealt{frederiks_2022}; \citealt{maan_2022}; \citealt{wang_2022a}; \citealt{tan_2026}; \citealt{xue_2026}). 
    This suggests that the physical processes responsible for FRBs are not limited to the radio band but involve broader energy release mechanisms.
    Currently, one of the leading scenarios is the magnetospheric model, in which coherent radio emission is produced by plasma generated through the release of magnetic energy in the magnetosphere of a magnetar (\citealt{kumar_2017}; \citealt{kumar_2020}; \citealt{lu_2020}).
    The temporal correlations of repeating FRBs exhibit characteristics similar to those of earthquakes (\citealt{totani_2023}; \citealt{tsuzuki_2024}; \citealt{yamasaki_2025}), and the intermittent energy injections that trigger FRBs may be related to activity in the neutron star's crust or to starquakes.    
    In this picture, rapid relaxation of external magnetic field lines can generate high-energy emission such as XRBs, and trapped fireballs are formed by pair plasma (e.g., \citealt{yamasaki_2022}). 
    Furthermore, X-ray photons escaping from the fireball are expected to be resonantly scattered by electrons, thereby producing an X-ray spectrum extending to high cutoff energies (\citealt{yang_2021}).

    Localization of repeating FRBs with subarcsecond precision has revealed that a subset of these sources are associated with persistent radio sources (PRSs).
    To date, PRSs have been robustly identified for four repeating FRBs: FRB 20121102A, FRB 20190417A, FRB 20190520B, and FRB 20240114A (\citealt{chatterjee_2017}; \citealt{marcote_2017}; \citealt{niu_2022}; \citealt{bruni_2025}; \citealt{moroianu_2025}; \citealt{zhang_2025b};).
    Observations of these PRSs have further constrained their sizes, indicating that they are compact regions with scales ranging from subparsec to at most a few tens of parsecs (\citealt{tendulkar_2017}; \citealt{niu_2022}; \citealt{moroianu_2025}; \citealt{zhang_2025b};  \citealt{bruni_2026}).
    These FRBs have been localized to host galaxies with known spectroscopic redshifts, and all of these hosts are star-forming galaxies (\citealt{tendulkar_2017}; \citealt{niu_2022}; \citealt{chen_2025}; \citealt{moroianu_2025}). 
    This suggests a possible link between ongoing star formation activity and the formation of FRB-PRS systems.
    In addition, a PRS candidate has been reported for FRB 20201124A, another active repeating FRB, although its physical size has not yet been constrained to the parsec scale (\citealt{ravi_2022}; \citealt{bruni_2024}).
    Furthermore, recent observations with uGMRT have identified PRS candidates associated with four additional FRBs (\citealt{pelliciari_2026a}; \citealt{pelliciari_2026b}; \citealt{pelliciari_2026c};).

    To explain the PRS emission, models have been proposed in which the PRS originates from synchrotron radiation in a magnetar wind nebula (MWN) formed by outflows from a central magnetar (\citealt{murase_2016}; \citealt{yang_2016}).
    Previous studies based on the expanding MWN-supernova remnant (SNR) model have shown that relatively young systems, with ages of $\lesssim 100$ yr, can reproduce the observed PRS fluxes and spectra (\citealt{margalit_2018}; \citealt{zhao_2021}; \citealt{bhattacharya_2026}; \citealt{zhao_2026}). 
    Moreover, these models can account for additional observational features, such as the decreasing dispersion measure (DM) and large rotation measure (RM) observed in sources like FRB 20121102A and FRB 20190520B (\citealt{niu_2022}; \citealt{snelders_2025}), making them a promising framework for understanding FRB-PRS systems.
    In addition to the expanding MWN-SNR scenario, binary models have also been discussed. In particular, the complex and highly variable RMs observed in FRB 20190520B and FRB 20201124A (\citealt{xu_2022}; \citealt{anna_2023}) have been suggested to originate from a binary system containing a magnetar and a massive stellar companion (\citealt{wang_2022b}; \citealt{rajwade_2023}; \citealt{zhao_2023};  \citealt{zhang_2025c}). Moreover, models involving young magnetars embedded in SNRs within massive binary systems have been proposed to account for both RM variability and long-term DM evolution \citep{wang_2025}.

    Among FRBs associated with PRSs, FRB 20240114A (\citealt{tian_2024}; \citealt{shin_2026}) stands out as the most active repeater discovered to date. 
    Observations by FAST between January and August 2024 (\citealt{zhang_2025a}) detected more than 10,000 bursts from this source, revealing dramatic temporal variations in its activity, with burst rates exceeding 700 per hour during its most active phases.
    This extreme level of activity has led to discussions on whether the total energy output can be sustained within the typical magnetic energy budget of a magnetar, although this issue remains unresolved.
    FRB 20240114A has been localized to a dwarf star-forming galaxy at a redshift of $z = 0.1306$, as revealed by optical spectroscopy (\citealt{chen_2025}).
    Also, polarization observations of individual bursts have provided measurements of DM, RM, and both linear and circular polarization fractions.
    While the DM remains relatively stable at $\sim 529\,\mathrm{pc\,cm^{-3}}$, the RM shows a gradual, approximately linear decline over several hundred days following an initial stable phase (\citealt{wang_2026}).
    This behavior suggests that the source resides in a dynamically evolving magneto-ionic environment.
    Recent full-Stokes, high-cadence observations of FRB 20240114A (\citealt{manaswini_2026}) have revealed that the polarization position angle (PPA) exhibits highly variable behaviors on timescales ranging from milliseconds to hours.
    The significant burst-to-burst variability in the PPA has been interpreted as possible evidence that FRB originates from multiple emission regions.

    The PRS associated with FRB 20240114A has been observed across multiple radio frequency bands between February and October 2024 (\citealt{bhusare_2025}; \citealt{bruni_2025}; \citealt{zhang_2025b}).
    In particular, observations at 1.5 GHz (\citealt{zhang_2025b}) suggest that the PRS flux exhibits variability (both rise and decline) on timescales of several months.
    After accounting for the effects of refractive scintillation in the Galactic interstellar medium (ISM), this month-scale flux variability has been argued to be intrinsic to the PRS itself (\citealt{zhang_2025b}).
    Furthermore, the variations in the PRS flux have been reported to be correlated with the activity of FRB 20240114A, suggesting that periods of enhanced, hyper-active FRB activity could influence the brightness of the PRS (\citealt{li_2026}; \citealt{zhang_2026}).

    In this study, we investigate the possibility that the PRS associated with FRB 20240114A is powered by synchrotron emission from relativistic electrons in a MWN.
    We propose that the flux variations suggested by recent PRS observations can be explained by additional electron injection into the MWN associated with episodes of hyper-active FRB activity driven by the central magnetar.
    Within this framework, we quantitatively examine how electrons supplied by hyper-active FRB activity affect the luminosity, spectrum, and temporal evolution of the PRS in an expanding MWN.
    By comparing the model predictions with the observed PRS flux variations, we assess whether the observed flux evolution can be reproduced and derive predictions for the future behavior of the PRS.
    We also discuss how modeling the PRS flux variability can provide constraints on the physical properties of the FRB emission source, including the intrinsic energy of FRBs and the bulk Lorentz factor of the FRB-emitting plasma.
    
    This paper is organized as follows.
    In Section \ref{sec:model}, we describe our model consisting of a magnetar embedded within an expanding MWN and supernova ejecta, and explain how the temporal evolution of the PRS luminosity and spectrum is calculated, as well as how the observed activity of FRB 20240114A is incorporated into the model.
    In Section \ref{sec:fitting}, we present the fitting procedure for the observed PRS flux variations and discuss the physical constraints on the FRB emission source derived from the fitting results.
    In Section \ref{sec:discussion}, we discuss implications for the intrinsic energetics of FRBs and the central magnetar, the future evolution of the PRS, and the validity of the magnetar scenario. 
    Finally, we summarize our conclusions in Section \ref{sec:summary}.

  \section{Modeling the PRS Light Curves and Spectra}
  \label{sec:model}
\begin{figure}
        \centering
        \includegraphics[width=1.0\linewidth]{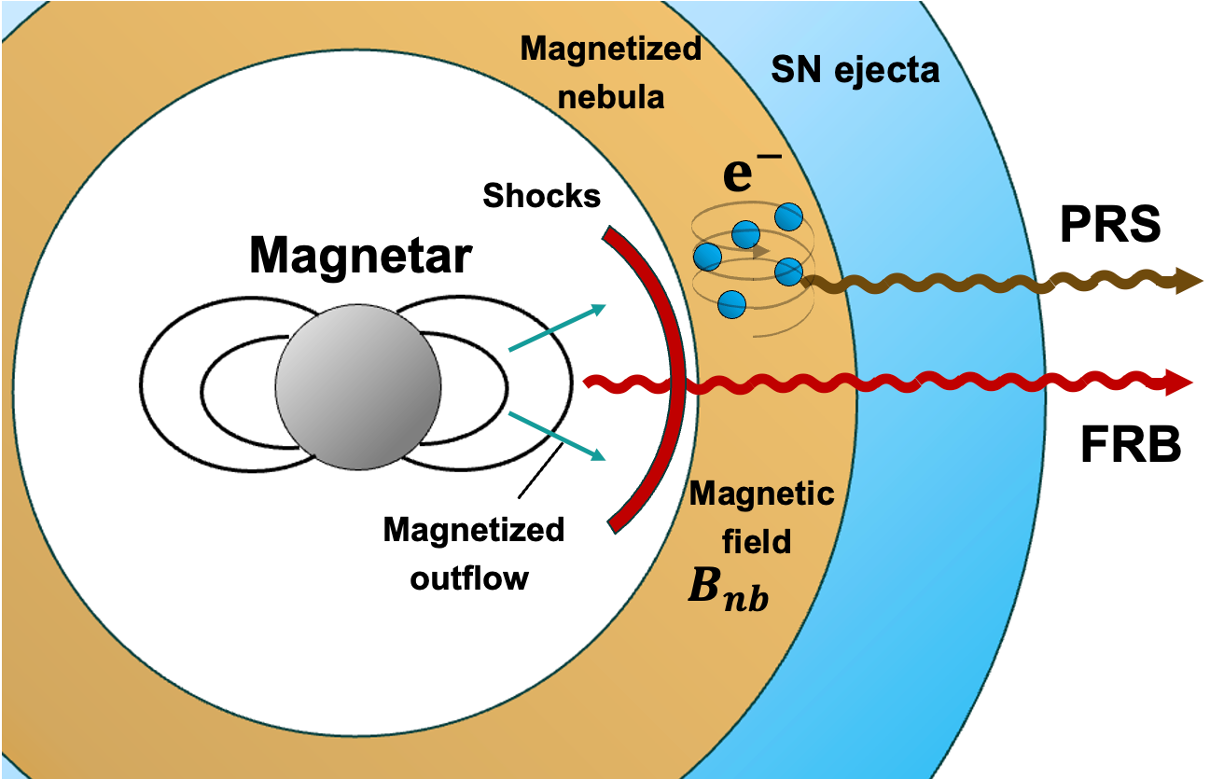}
        \caption{A schematic picture of a young magnetar surrounded by a magnetar wind nebula (MWN) and supernova ejecta. The PRS emission associated with the FRB is produced by synchrotron radiation from relativistic electrons injected into the MWN. The particles and magnetic energy are assumed to be injected into the MWN predominantly through magnetic flares from the magnetar.}
        \label{fig:schematic_picture}
      \end{figure}

    We consider that the PRS emission is produced by synchrotron radiation from relativistic plasma injected from a magnetar into the MWN.
    Figure \ref{fig:schematic_picture} shows a schematic picture of our model, in which a young magnetar is surrounded by a MWN that is embedded within supernova ejecta.
    Here, the MWN refers to the relativistic bubble inflated by energy injection from the magnetar.
    In our numerical calculations, we solve the coupled dynamical evolution of the MWN and the surrounding supernova ejecta, and compute the temporal evolution of the radio light curve and spectrum of the MWN by solving the electron transport equation.
    In particular, we investigate how plasma injection associated with the hyper-active phase of FRB 20240114A affects the radio light curve of the PRS, and attempt to explain the observed flux variability of the PRS.

   \subsection{Energy Injection from the Magnetar}
   \label{subsec:magnetar_energy_injection}
      In modeling the energy injection from the magnetar into the MWN, we assume that the PRS is predominantly powered by magnetic energy dissipation in flares near the surface of the magnetar, and that the FRBs are also generated by magnetospheric activity of the magnetar.
      Therefore, although the magnetar can also inject energy via spin-down powered outflows driven by unipolar induction, we treat the spin-down luminosity as a subdominant contribution to the PRS brightness and hence ignore it.

      Energy injection via magnetic flares has been discussed in previous studies as a mechanism to explain the PRS associated with active repeating FRBs, such as FRB 20121102A and FRB 20190520B (\citealt{beloborodov_2017}; \citealt{margalit_2018}; \citealt{zhao_2021}).
      \citealt{margalit_2018} modeled the energy injection rate from magnetic flares as a power-law decay

      \begin{equation}
        L_{\rm mag}(t) = (\alpha - 1) \frac{E_{\rm mag,i}}{t_{\rm inj}} \left( \frac{t}{t_{\rm inj}} \right)^{-\alpha}, \, t\geq  t_{\rm inj}, \, \alpha \geq  1,
        \label{magnetic_flare_luminosity}
      \end{equation}

      \noindent where $E_{\rm mag,i}=B_{\rm int}^2 R_{\ast}^3/6 \sim 3 \times 10^{47}  \ B_{\rm int,15}^2 \ {\rm erg}$ is the initial magnetic energy of the magnetar with an initial magnetic field strength of $B_{\rm int} = 10^{15} B_{\rm int,15}$ G, and $t_{\rm inj}$ is the time when the energy injection from magnetic flares starts.
      Although magnetic flares occur sporadically, this smooth function represents the mean level of magnetospheric activity that gradually decays over time.
      In this study, we adopt the parameters $t_{\rm inj} \approx  0.6 \ \mathrm{yr}$ and $\alpha \approx  1.3$ used in a fiducial model for the fitting to the PRS associated with FRB 20121102A (\citealt{margalit_2018}).
      Although there are substantial uncertainties in the value of $t_{\rm inj}$ and the decay index $\alpha$, we assume that FRB 20240114A undergoes active repeating activity driven by a physical mechanism similar to that of FRB 20121102A, and therefore employ these values in our model.
      The impact of these parameters on the conclusions of this study is briefly discussed in Section \ref{subsec:magnetar_energy_budget}.

    \subsection{Energy Injection from the Activity of FRB 20240114A}
    \label{subsec:frb_energy_injection}
      To investigate how the extremely active FRB activity of FRB 20240114A affects the PRS, we introduce an additional contribution to the injected energy associated with the transition of FRB 20240114A into a hyper-active phase, in addition to the smoothly evolving energy injection of Equation (\ref{magnetic_flare_luminosity}). 
      In order to calculate the contribution from FRB 20240114A, we use the observational data of FRB 20240114A reported by \citealt{zhang_2025a}, which monitored FRB activity with FAST in the L-band ($1$--$1.5 \, \rm{GHz}$) over a period of approximately seven months starting from 2024 January 28.
    
      As described by \citealt{zhang_2025a}, the burst rate of FRB 20240114A appears to exhibit two distinct episodes with a primary peak on 12 March (UT) and a secondary peak on 28 July (UT).
      Therefore, we divide the time series of the observational data into four episodes (Table \ref{tab1:episodes}): two episodes corresponding to clearly enhanced activities and the remaining two periods. 
      
      The beaming-corrected radio burst energy for each burst is 
      \begin{equation}
        \begin{split}
          E_{\rm FRBr} &= b_{\rm FRB} \cdot E_{\rm FRBr,iso} \\
          E_{\rm FRBr,iso} &= \frac{4 \pi D_{\rm L}^2}{1+z} \, F_{\nu,{\rm obs}} \, \Delta \nu_{\rm obs}
        \end{split}
        \label{E_FRBr}
      \end{equation}
      \noindent where $E_{\rm FRBr, iso}$ is the isotropic-equivalent radio burst energy, and $b_{\rm FRB}$ ($0 \leq b_{\rm FRB} \leq 1$) is the global beaming factor, accounting for the possibility that the true emitted energy is smaller than the isotropic-equivalent energy. 
      Here, $z$ is the redshift of FRB 20240114A, $D_{\rm L}$ is the luminosity distance, $F_{\nu,{\rm obs}}$ is the observed fluence of the burst, and $\Delta \nu_{\rm obs}$ is the observed bandwidth.
      While the isotropic energy of an FRB depends on its spectral shape, repeating bursts typically exhibit narrow spectra within the observing band, with the central frequency and bandwidth specific to each burst \citep{zhang_2023b}. 
      Therefore, $F_{\nu,{\rm obs}}$ is defined as the average burst fluence over the observed frequency range.
      The luminosity distance $D_{\rm L} = 633.87 \ {\rm Mpc}$ at the spectroscopic redshift of $z=0.1306$ (\citealt{chen_2025}) is calculated based on the standard Planck cosmological parameters (\citealt{planck_2016}).
      
\begin{figure}[htbp]
          \centering
          \includegraphics[width=1.0\linewidth]{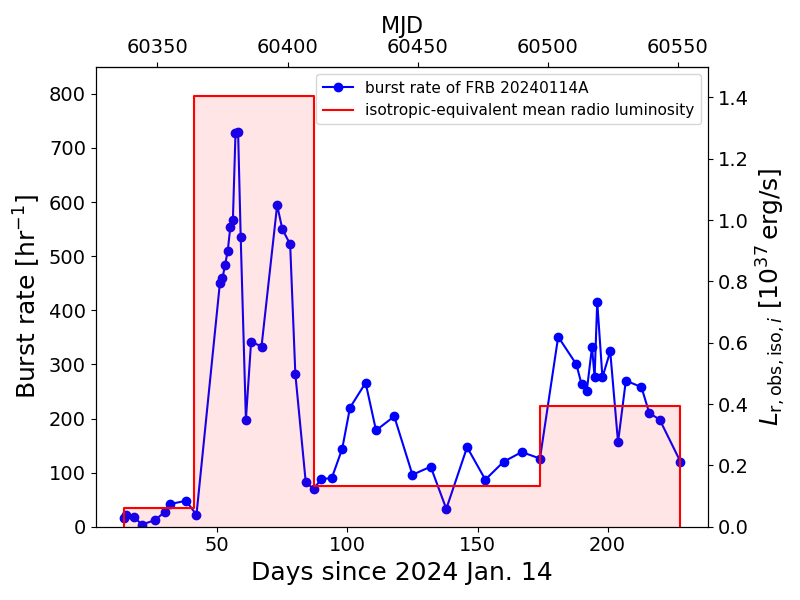}
          \caption{The time series of the burst rate of FRB 20240114A, overlaid with the isotropic-equivalent mean radio luminosity $L_{{\rm r,obs,iso},i}$ for each of the four divided episodes. The blue points represent the burst rate observed by FAST (\citealt{zhang_2025a}), 
          while the red line represents the values of $L_{{\rm r,obs,iso},i}$ for each episode calculated from Equation (\ref{L_r,obs,i}). }
          \label{fig:burst_rate}
      \end{figure}

      Then we define the beaming-corrected mean radio luminosity for the $i$-th episode $(i=1,2,3,4)$ as
      \begin{equation}
        \begin{split}
          L_{{\rm r,obs},i} &= b_{\rm FRB} \cdot L_{{\rm r,obs,iso},i} \\
          L_{{\rm r,obs,iso},i}  &= \frac{1}{\Delta t_i} \cdot \zeta_{i}^{-1} \sum_{j=1}^{N_i} E_{{\rm FRBr,iso},j},
        \end{split}
       \label{L_r,obs,i}
      \end{equation}
      where $L_{{\rm r,obs,iso},i}$ is the isotropic-equivalent mean radio luminosity.
      Here, $E_{{\rm FRBr,iso},j}$ is the isotropic-equivalent radio energy of the $j$-th burst in the $i$-th episode, $N_i$ is the number of detected bursts in that episode, $\zeta_i$ is the duty cycle of observation during the episode, and $\Delta t_i$ is the full duration of the episode.
      Here, $\zeta_i$ is calculated as the ratio of the total observation time to the full duration of the $i$-th episode.
      The full duration and the number of bursts for each episode, along with the corresponding values of $\zeta_i$ and $L_{\rm r,obs,iso,i}$ calculated from Equation (\ref{L_r,obs,i}), are summarized in Table \ref{tab1:episodes}.
      In addition, Figure \ref{fig:burst_rate} shows the time series of the burst rate, overlaid with the values of $L_{{\rm r,obs,iso},i}$ for each episode.

      \renewcommand{\arraystretch}{1.5}
      \begin{table}[htbp]
        \caption{Radio properties of FRB 20240114A for each of the four divided episodes}
        \label{tab1:episodes}
        \resizebox{\columnwidth}{!}{
          \Huge
          \hspace{-5cm}
          \begin{tabular}{cccccc}
            \hline
            Episode &
            Start MJD &
            Full duration [days] &
            $N_i$ &
            $\zeta_i$ &
            $L_{{\rm r,obs,iso},i}$ [erg s$^{-1}$] \\
            \hline
            1st &60337.229849817 &27.92  &106  &0.0067  &6.3 $\times$ 10$^{35}$  \\
            2nd &60365.152766484 &44.89  &8772  &0.015  &1.4 $\times$ 10$^{37}$  \\
            3rd &60410.041655373 &87.83  &881  &0.0029  &1.3 $\times$ 10$^{36}$  \\
            4th &60497.872210928 &53.84  &1794  &0.0051  &3.9 $\times$ 10$^{36}$  \\
            \hline
          \end{tabular}
        }
      \end{table}

    Furthermore, to convert this into the actual energy input, we consider two additional factors: FRBs that are not observed off-axis, and the fraction of the energy that is converted into radio emission.
      The mean radio luminosity including the contribution from bursts that are missed owing to off-axis viewing is given by
        \begin{equation}
          L_{{\rm radio},i} = f_{\rm on-axis}^{-1} \cdot L_{{\rm r,obs},i},
          \label{L_radio,i}
        \end{equation}
       where $f_{\rm on-axis}$ denotes the fraction of all FRBs that are observed on-axis.
      Next, taking into account 
      the energy that goes into radiation at other wavelengths and the kinetic energy of the outflow, the total energy injection rate in the $i$-th episode is given by
        \begin{equation}
          L_{{\rm FRBt},i} = \eta_{\rm r}^{-1} \cdot L_{{\rm radio},i},
          \label{L_FRBt,i}
        \end{equation}
      where the parameter $\eta_{\rm r}$ denotes the radio radiation efficiency.
      Although there is an observational upper limit of $\eta_{\rm r} = 10^{-4}$--$10^{-5}$, inferred from FRB 20200428D associated with X-ray bursts (\citealt{mereghetti_2020}; \citealt{li_2021}), the value of $\eta_{\rm r}$ is highly uncertain and may vary significantly among different FRBs.
      Combining Eqs. (\ref{L_r,obs,i}), (\ref{L_radio,i}) and (\ref{L_FRBt,i}), the total energy injection rate in the $i$-th episode can be expressed in terms of $L_{{\rm r,obs,iso},i}$ inferred directly from the observed burst sample, as
         \begin{equation}
           L_{{\rm FRBt},i} = b_{\rm FRB} \cdot f_{\rm on-axis}^{-1} \cdot \eta_{\rm r}^{-1} \cdot L_{{\rm r,obs,iso},i}.
         \end{equation}
      For convenience, we define $f_{\rm FRBt} \equiv b_{\rm FRB} \cdot f_{\rm on-axis}^{-1} \cdot \eta_{\rm r}^{-1}$ so that 
         \begin{equation}
           L_{{\rm FRBt},i} = f_{\rm FRBt} \cdot L_{{\rm r,obs,iso},i}.
           \label{L_FRBt,i}
         \end{equation}
      We treat $f_{\rm FRBt}$ as a free parameter in our model, and explore how the parameter affects the PRS light curve and spectrum.
      Hereafter, we refer to the energy injection according to Equation (\ref{L_FRBt,i}) from FRB 20240114A divided into four episodes as the "\textit{four-episode FRB injection}".

    \subsection{Dynamical Evolution of the MWN and Supernova Ejecta}
    \label{subsec:MWN_dynamics}
      We compute the time evolution of the MWN driven by the injection of magnetic energy from the magnetar as follows.
      We assume that the energy injected by the magnetar is deposited into a bubble of relativistic plasma and magnetic fields surrounding the magnetar.
      The bubble drives a shock into the surrounding supernova ejecta gas and sweeps it up. The interior of the bubble is assumed to be a uniform, isotropic relativistic fluid with an adiabatic index $\gamma = 4/3$ (\citealt{ostriker_1971}; \citealt{chevalier_1977}).
      In this work, the shocked ejecta is treated by the thin-shell approximation, in which it is approximated as a thin shell enclosing the relativistic bubble (see e.g., \citealt{chevalier_1992}).
      Although synchrotron radiation may also be produced at the shock driven into the supernova ejecta, the energy dissipated at this shock is expected to be substantially smaller than that dissipated at the termination shock within the bubble (e.g., \citealt{chevalier_1992}). Consequently, we assume that the observed non-thermal emission is dominated by relativistic particles in the bubble accelerated by the wind termination shock.

      The evolution of radius $R_{\rm nb}$ and velocity $V_{\rm nb}$ of the thin shell with mass $M$ is described by (\citealt{ostriker_1971}; \citealt{chevalier_1977}; \citealt{chevalier_1992})
        \begin{eqnarray}
          M \frac{dV_{\rm nb}}{dt} = 4 \pi R_{\rm nb}^2 \ [p_i - \rho_{\rm ej}(V_{\rm nb} - V_{\rm ej})^2] \\
          \frac{d}{dt} (4 \pi R_{\rm nb}^3 \ p_i) = L(t) - 4 \pi R_{\rm nb}^2 \ p_i \frac{dR_{\rm nb}}{dt},
        \end{eqnarray}
      where $p_i$ is the bubble pressure at the inner boundary of the shell, and $\rho_{\rm ej}$ and $V_{\rm ej}$ are the density and velocity of the supernova ejecta immediately outside the shell, respectively.
      The thin-shell mass $M$ represents the ejecta mass swept up within the MWN radius $R_{\rm nb}$, and is given by $M = \int_0^{R_{\rm nb}} 4 \pi r^2 \rho_{\rm ej} \, dr$.
      The total energy injection rate from the magnetar, $L(t)$, is given by Eqs. (\ref{magnetic_flare_luminosity}) and (\ref{L_FRBt,i}) as
        \begin{equation}
          \begin{split}
          L(t) = 
            \begin{cases}
              & L_{\rm mag}(t), \\
              & (t \geq t_{\rm inj}, \text{outside the four burst episodes}) \\
              & \\
              & L_{\rm mag}(t) + L_{\rm FRBt}  (t), \\
              & (\text{during the four burst episodes}).
            \end{cases}
          \end{split}
        \end{equation}
      Details of how we determine the onset time for the four-episode FRB injection will be presented in Section \ref{subsec:fitting_procedure}. 
      Since we consider the early evolutionary stage of the MWN, we assume free expansion of the supernova ejecta.
      Accordingly, we take $V_{\rm ej}$ to be $r/t$.
      The characteristic expansion velocity of the ejecta scales with the supernova explosion energy and the ejecta mass, and details of the modeling are described in Section \ref{subsec:ejecta_modeling}.

      By solving the nebular dynamics we get $R_{\rm nb}(t)$ and $p_i(t)$, the time-evolving average magnetic field $B_{\rm nb}$ in the nebula can be determined, which is given by
        \begin{equation}
          B_{\rm nb}(t) \approx \sqrt{\frac{6 \epsilon_B}{R_{\rm nb}^3(t)} \int_{t_{\rm inj}}^t L(t') dt'},
          \label{average_magnetic_field}
        \end{equation}
      where $\epsilon_B$ is the magnetic field amplification efficiency, which is fixed to $\epsilon_B = 0.01$ in this work.
      This value is motivated by spectral modelings of the Crab Nebula and several other young pulsar wind nebulae (PWNe) in previous studies, where multi-wavelength observations are well fitted with $\epsilon_B$ of order $10^{-2}$ (\citealt{torres_2013}; \citealt{zhu_2023}).
      However, substantial uncertainty remains in the value of $\epsilon_B$, and variations of up to an order of magnitude cannot be excluded. 
      We briefly discuss the impact of this uncertainty on our conclusions in Section \ref{subsec:magnetar_energy_budget}.
      Equation (\ref{average_magnetic_field}), in which the average magnetic field is estimated from the time integral of $L(t)$, has been commonly adopted in previous studies instead of using the thermal energy stored in the relativistic bubble (\citealt{tanaka_2010}; \citealt{murase_2015}; \citealt{murase_2016}).

    \subsection{Modeling the Supernova Ejecta Structure}
    \label{subsec:ejecta_modeling}
      The density structure of a freely expanding supernova ejecta can be approximated by an inner relatively flat profile $\rho_{\rm ej, i} = A r^{-m} t^{m-3} \ (m=0$--$1)$ and an outer steep profile $\rho_{\rm ej, o} = B r^{-n} t^{n-3} \ (n \approx 9)$, which are assumed to be continuous (\citealt{chevalier_1989}).
      The two parts of the density profile intersect at a velocity given by
        \begin{equation}
          V_{t} = \left[ \frac{2(5-m)(n-5)}{(3-m)(n-3)} \frac{E_{\rm sn}}{M_{\rm ej}} \right]^{1/2},
        \end{equation}
      where $E_{\rm sn}$ is the total energy of the supernova explosion and $M_{\rm ej}$ is the ejecta mass.
      We define $R_{\rm ej}(t) = V_{t} \, t$ as the boundary between the inner and outer density profiles of the supernova ejecta.
      We consider only the case where the thin shell is still within the inner profile, and hence  $\rho_{\rm ej} = \rho_{\rm ej, i}$.
      We obtain the normalization constant $A$ of the $\rho_{\rm ej, i}$ as
        \begin{eqnarray}
            A &=& \left(\frac{n-3}{n-m}\right) 
            \left[\frac{(3-m)(n-3)}{2(5-m)(n-5)} 
            \frac{M_{\rm ej}}{E_{\rm sn}}\right]^{(3-m)/2} \nonumber \\ 
            && \times \, \frac{(3-m) \, M_{\rm ej}}{4\pi}.
        \end{eqnarray}
      In this work, we set $m=0$, $n=9$ (\citealt{chevalier_1992}).

      With regard to the magnetar progenitor, we consider two types of supernova explosions: a typical core-collapse supernova (CCSN) and an ultra-stripped supernova (USSN).
      Although CCSNe exhibit a range of ejecta masses depending on the degree of envelope stripping prior to explosion, observational and theoretical studies indicate that typical CCSNe have ejecta masses of a few to $\sim 10\,M_\odot$ and explosion energies of order $10^{51}\,\mathrm{erg}$ (\citealt{dessart_2016}; \citealt{prentice_2019}; \citealt{burrows_2021}).
      In this work, we adopt representative values of $M_{\rm ej} = 3.0\,M_\odot$ and $E_{\rm sn} = 10^{51}\,\mathrm{erg}$, adopted in previous modeling of PRSs (\citealt{bhattacharya_2026}).
      In contrast, USSNe represent a relatively rare and distinct class of explosions arising from progenitors that have lost most of their outer envelopes through close binary interaction. We adopt typical values of $M_{\rm ej} = 0.1\,M_\odot$ and $E_{\rm sn} = 10^{50}\,\mathrm{erg}$, as estimated from both theoretical models and observationally inferred events (\citealt{tauris_2015}; \citealt{de_2018}).

    \subsection{Free-Free Absorption Constraints from the FRB Detection}
    \label{subsec:free_free_absorption}
      FRBs produced in the magnetar magnetosphere can be attenuated by scattering and absorption processes occurring in the MWN and the surrounding supernova ejecta. Among these, free-free absorption in the supernova ejecta is the most relevant process (see e.g., \citealt{murase_2017}).
      The magnetar should be old enough so that the free-free optical depth $\tau_{\rm ff}$ at $\sim$ GHz frequencies of the supernova ejecta is less than unity at the time of FRB emission;
        \begin{equation}
          \left. \tau_{\rm {ff}} \right|_{\nu = 1 \, {\rm GHz}} \approx 2.1 \times 10^{-25} \ \mathcal{T}_{\rm e,4}^{-1.35} \int dr \ n_{\rm {e,ej}} n_{\rm {i,ej}} \bar{Z}^2 \lesssim 1,
        \end{equation}
      where $\mathcal {T}_{\rm e,4}$ is the electron temperature normalized to $10^4\,{\rm K}$, $n_{\rm {e,ej}}$ and $n_{\rm {i,ej}}$ are the number densities of free electrons and ions in the supernova ejecta, respectively, and $\bar{Z}$ is the average ionic charge number.
      The number density of free electrons in the supernova ejecta is given by $n_{\rm {e,ej}} \approx 3 M_{\rm ej} / (4 \pi R_{\rm ej}^3 \mu_e \bar{A} m_{\rm H})$, where $\bar{A}$ denotes the mean atomic mass number, which we take to be $\bar{A} = 10$, $m_{\rm H}$ is the mass of a hydrogen atom, and we set $\bar{Z} \sim \bar{A}/2$ for simplicity.
      For the ionization state, we assume singly ionized material $(\mu_e \approx 1)$, and accordingly set the electron temperature as $\mathcal{T}_e \sim 10^4 \ \rm K$ (see \citealt{kashiyama_2017}).

    \subsection{Electron Transport and Synchrotron Emission from the MWN}
    \label{subsec:electron_transport}
      We calculate the time evolution of the electron energy distribution in the MWN by solving the electron transport equation, which accounts for the injection of relativistic electrons from the magnetar, as well as cooling processes such as synchrotron radiation and adiabatic expansion.
      The synchrotron emission from the MWN is then computed based on the electron energy distribution and the magnetic field strength in the nebula.

      Assuming a homogeneous and isotropic particle distribution in the MWN, the distribution function can be volume-integrated to yield the electron energy distribution $N_e(\gamma, t)$, defined such that $N_e(\gamma,t)\,d\gamma$ gives the number of electrons within the Lorentz-factor interval $[\gamma,\gamma+d\gamma]$.
      The evolution of $N_e(\gamma, t)$ is described by the continuity equation in the energy space,
        \begin{equation}
          \frac{\partial}{\partial t} N_e(\gamma, t) + \frac{\partial}{\partial \gamma} [\dot{\gamma}(\gamma, t) N_e(\gamma, t)] = \dot{Q}_{\rm inj}(\gamma, t).
        \end{equation}

      The electron injection term $\dot{Q}_{\rm inj}(\gamma, t)$ is determined by the energy injection from the magnetar, and the acceleration process inside or around the termination shock before the electrons enter the MWN.
      We assume that the relativistic electron-positron pair outflow driven by magnetic energy release from the magnetar is efficiently dissipated near the termination shock and/or within the nebula, and that a substantial fraction of its energy is converted into the downstream non-thermal particle distribution. It is also assumed that the injected electrons are accelerated to a power-law distribution in energy, which is expected from the diffusive shock acceleration process (\citealt{bell_1978}; \citealt{blandford_1978}), and that the minimum Lorentz factor of the injected electrons is determined by the bulk Lorentz factor $\gamma_{\rm b}$ of the outflow from the magnetar.
      Thus, the injection rate is given by
        \begin{equation}
          \gamma^2 \dot{Q}_{\rm inj}(\gamma, t) = \frac{\epsilon_e L(t)}{m_e c^2 \mathcal{R}_0} \left(\frac{\gamma}{\gamma_{\rm b}}\right)^{2-q}, \, \gamma \geq \gamma_{\rm b}, 
          \label{injection_rate}
        \end{equation}
      where $\epsilon_e \approx 1-\epsilon_B$ is the injection efficiency, $m_e$ is the electron mass, and $c$ is the speed of light.
      We fix the power-law index to $q=2.5$ derived from spectral fitting of the Crab Nebula (\citealt{tanaka_2010}), and
      $\mathcal{R}_0$ is the normalization factor, which is given by $\mathcal{R}_0 \sim (q-2)^{-1} \sim 2$.
      In previous studies, the injection term has often been modeled using a broken power-law distribution (see e.g., \citealt{tanaka_2010}; \citealt{murase_2015}). However, the actual injection spectrum of electrons is subject to substantial uncertainty. Therefore, in this work, we minimize assumptions by adopting only the single power-law index associated with the acceleration process, while treating the electron distribution at Lorentz factors below $\gamma_{\rm b}$ as being formed through cooling processes.
      The value of $\gamma_{\rm b}$ is determined from the fitting of the PRS light curves, as described in the following section.

      The cooling term $\dot{\gamma}(\gamma, t)$ for electrons accounts for radiative losses due to synchrotron emission and inverse Compton scattering, together with adiabatic expansion, i.e., 
        \begin{equation}
          \dot{\gamma}(\gamma, t) = \dot{\gamma}_{\rm syn}(\gamma, t) + \dot{\gamma}_{\rm IC}(\gamma, t) + \dot{\gamma}_{\rm ad}(\gamma, t).
        \end{equation}
      The electron cooling due to inverse Compton scattering off the cosmic microwave background (CMB) is not expected to be important compared to synchrotron and adiabatic cooling.
      Next, we evaluate the time evolution of the relative importance of synchrotron self-Compton (SSC) and synchrotron cooling.
      The ratio of the SSC cooling rate to the synchrotron cooling rate can be estimated as $\dot{\gamma}_{\rm SSC} / \dot{\gamma}_{\rm syn} \sim U_{\rm syn} / U_B$, where $U_{\rm syn}$ is the energy density of synchrotron photons and $U_B$ is the magnetic energy density in the MWN.
      The synchrotron photon energy density can be approximated as $U_{\rm syn} \sim P_{\rm syn} / (4 \pi R_{\rm nb}^2 c)$, where $P_{\rm syn}(t)$ is the power of the synchrotron radiation: 
        \begin{equation}
          P_{\rm syn}(t) \sim \frac{4}{3} \sigma_{\rm T} c \gamma_{\rm b}^2 U_B(t) \cdot \gamma_{\rm b} N_e(\gamma_{\rm b}, t),
        \end{equation}
      where $\sigma_{\rm T}$ is the Thomson cross section, and $\gamma_{\rm b} N_e(\gamma_{\rm b}, t)$ represents the number of the electrons around $\gamma = \gamma_{\rm b}$.
      If we define the cumulative injected energy as $E_{\rm inj}(t) = \int_{t_{\rm inj}}^t L(t') dt'$, the injected particle energy can be approximated as $N_e(\gamma_{\rm b}, t) \gamma_{\rm b}^2 m_e c^2 \sim \epsilon_e E_{\rm inj}(t)$.
      Therefore, the ratio $U_{\rm syn} / U_B$ can be estimated as
        \begin{eqnarray}
          \frac{U_{\rm syn}}{U_B} &\sim& \frac{P_{\rm syn}(t)}{4 \pi R_{\rm nb}^2(t) c U_B(t)} \nonumber \\
          &\sim& \frac{\sigma_{\rm T} \gamma_{\rm b} \epsilon_e E_{\rm inj}(t)}{3 \pi m_e c^2 R_{\rm nb}^2(t)} .
          \label{ssc_syn_ratio}
        \end{eqnarray}
      The cumulative injected energy is given by $E_{\rm inj}(t) = E_{\rm mag,i}(1-(t_{\rm inj}/t)^{\alpha-1})$ when the four-episode FRB injection is not included.
      For the MWN radius $R_{\rm nb}(t)$, when the energy injection rate follows a power-law decline $L \propto t^{-\alpha}$ with $\alpha > 1$, the expansion of the young MWN is expected to proceed at nearly constant velocity (\citealt{chevalier_1992}; \citealt{bucciantini_2004}; \citealt{gelfand_2009}).
      Therefore, from Equation (\ref{ssc_syn_ratio}), the ratio $U_{\rm syn} / U_B$ approximately evolves as $U_{\rm syn} / U_B \propto t^{-2}$ for $t > t_{\rm inj}$.
      The cooling via SSC scattering can be important at early times ($\sim t_{\rm inj}$) below the self-absorption frequency, but rapidly becomes negligible as the MWN expands (see e.g., \citealt{tanaka_2010}; \citealt{margalit_2018}).
      In this study, we focus on investigating how the activity of FRB 20240114A affects the radiation from the MWN, with our modeling primarily concerned with epochs after FRBs are able to escape from the MWN and the supernova ejecta.
      Therefore, to save the computational cost, we do not include the inverse Compton cooling term in our calculations.

      The synchrotron cooling and the adiabatic cooling are respectively given by (see e.g., \citealt{rybicki_1979})
        \begin{equation}
          \begin{split}
            \dot{\gamma}_{\rm syn}(\gamma, t) &= -\frac{\sigma_{\rm T} B_{\rm nb}^2(t)}{6 \pi m_e c} \gamma^2 \\
            \dot{\gamma}_{\rm ad}(\gamma, t) &= -\frac{V_{\rm nb}(t)}{R_{\rm nb}(t)} \gamma.
          \end{split}
        \end{equation}
      The synchrotron emission from the MWN is calculated based on the electron energy distribution and the average magnetic field strength in the nebula.
      The synchrotron radiation luminosity $L_{\nu, {\rm syn}}(t)$ at a given frequency $\nu$ and time $t$ is given by
        \begin{equation}
          L_{\nu, {\rm syn}}(t) = 4 \pi^2 R_{\rm nb}^2(t) \frac{j_{\nu}}{\alpha_{\nu}} \left(1-e^{-\alpha_{\nu} R_{\rm nb}(t)}\right),
        \end{equation}
      derived from the synchrotron emissivity and absorption coefficient
        \begin{equation}
          \begin{split}
            j_{\nu} &= \int \frac{N_e(\gamma, t) P(\nu, \gamma, t)}{4 \pi} d\gamma \\
            \alpha_{\nu} &= -\int \frac{\gamma^2 P(\nu, \gamma, t)}{8 \pi m_e \nu^2} \frac{\partial}{\partial \gamma} \left[ \frac{N_e(\gamma, t)}{\gamma^2} \right] d\gamma,
          \end{split}
        \end{equation}
      respectively, where 
        \begin{equation}
          \begin{split}
            P(\nu, \gamma, t) &= \frac{\sqrt{3} e^3 B_{\rm nb}(t)}{m_e c^2} F\left( \frac{\nu}{\nu_c(t)} \right) \\
            \nu_c(t) &= \frac{3 e B_{\rm nb}(t)}{4 \pi m_e c} \gamma^2.
          \end{split}
        \end{equation}
      Here, $F(x)$ is defined as $F(x) \equiv x \int_x^{\infty} K_{5/3}(y) dy$, where $K_{5/3}(y)$ is the modified Bessel function of order $5/3$.
      The observed flux density of the PRS is then readily obtained as $F_{\nu} = (1+z) L_{\nu, {\rm syn}} / (4 \pi D_{\rm L}^2)$.
      The typical Lorentz factor of electrons responsible for the MWN radio emission increases with time, and crosses at some point the peak Lorentz factor of the electron SED, leading to a rise and subsequent decline of the radio light curve (see e.g., \citealt{margalit_2018}).

  \section{Fitting the PRS Light Curves and Spectra to Observational Data}
  \label{sec:fitting}
    Here we compare the PRS light curves and SED calculated by our modeling with the observational data of the PRS associated with FRB 20240114A.
    This comparison enables us to place constraints on the physical parameters related to the FRB emission properties and the MWN evolution.

    \subsection{Fitting Procedure and Parameter Estimation}
    \label{subsec:fitting_procedure}
\begin{figure}[htbp]
        \centering
        \includegraphics[width=1.0\linewidth]{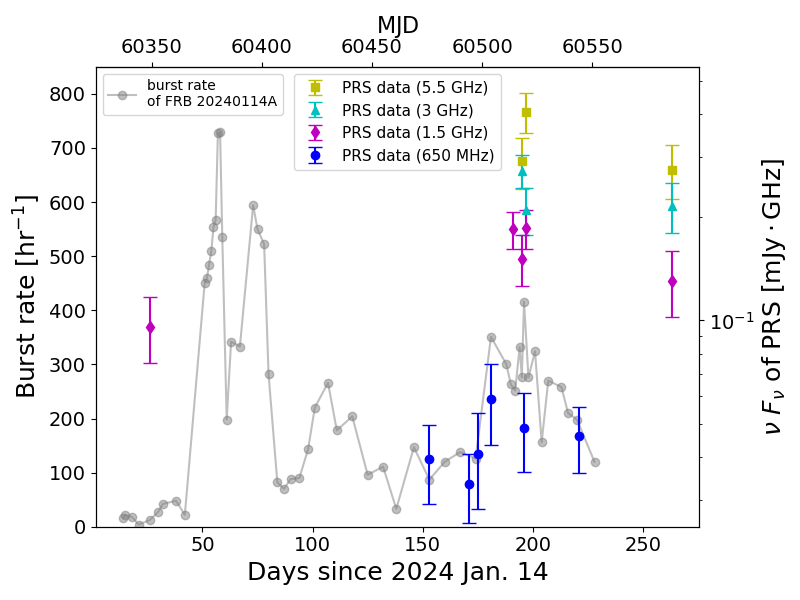}
        \caption{The time series of the burst rate (light gray points) of FRB 20240114A observed with FAST (\citealt{zhang_2025a}), overlaid with the observed $\nu F_{\nu}$ data of the PRS at 1.5, 3, and 5.5 GHz obtained with the VLA and MeerKAT between February and October 2024 (\citealt{zhang_2025b}), together with 650 MHz observations conducted with the GMRT between June and August 2024 (\citealt{bhusare_2025}). 
        The error bars on the PRS data indicate the 1$\sigma$ observational uncertainties.}
        \label{fig:frb_prs_data}
      \end{figure}

      The data used for the fitting consist of observations at 1.5, 3, and 5.5 GHz obtained with the VLA and MeerKAT between February and October 2024 (\citealt{zhang_2025b}), together with 650 MHz observations conducted with the GMRT between June and August 2024 (\citealt{bhusare_2025}).
      Figure \ref{fig:frb_prs_data} shows an overlay of the temporal evolution of the burst rate of FRB 20240114A and the observational data of the PRS. The PRS spectrum was observed in late July 2024, showing an indication for a synchrotron self-absorption (SSA) break at around 1.6 GHz (\citealt{zhang_2025b}).
      
      The model parameter of primary focus in this study is $f_{\rm FRBt}$, which relates the energy injected into the MWN to the observed FRB activity.
      Other fitting parameters are the magnetar initial magnetic field strength $B_{\rm int}$, the age $t_{\rm age}$ of the FRB-PRS system corresponding to the epoch at which the PRS data were obtained, and the bulk Lorentz factor $\gamma_{\rm b}$ of the electrons responsible for the FRB emission.
      As shown in Figure \ref{fig:frb_prs_data}, the first 1.5 GHz observation was obtained during a period when the FRB burst activity was at its lowest level, at which the contribution from the four-episode FRB injection is minimal.
      Thus, we define $t_{\rm age}$ as the time corresponding to this first 1.5 GHz data point, which was obtained at MJD 60349.
      
      Therefore, $B_{\rm int}$ and $t_{\rm age}$, which determine the average energy release history $L_{\rm mag}(t)$ of the magnetar prior to the four-episode FRB injection, are constrained by the first 1.5 GHz data point, while $f_{\rm FRBt}$ is constrained by the temporal variation behavior of the subsequent data points.   In the magnetically powered MWN scenario, $\gamma_{\rm b} \sim 10^3$ has been adopted in recent phenomenological modelling of FRB-associated PRSs (see e.g., \citealt{bhattacharya_2026}). Motivated by this, we explore a broader range of $\gamma_{\rm b}$, varying it $10$--$10^4$.
      It should be noted that the value of $B_{\rm int}$ scales with $t_{\rm inj}$ and $\alpha$ (see Equation \ref{magnetic_flare_luminosity}).
      As discussed in Section \ref{subsec:magnetar_energy_injection}, $t_{\rm inj}$ and $\alpha$ are subject to large uncertainties, and therefore the true value of $B_{\rm int}$ cannot be tightly constrained in the present fitting.

      Since a rigorous statistical fitting using methods such as MCMC is computationally too expensive, we instead estimate the approximate parameter ranges by the following two steps. As the first step, we fix $\gamma_{\rm b}$ to an arbitrary value and compute the time evolution of the PRS SED without including the four-episode FRB injection.
      We then determine the two parameters, $B_{\rm int}$ and $t_{\rm age}$, so that the model 
      can reproduce both the SSA break observed in late July 2024 and the radio flux measured in the first 1.5 GHz observation. 
      In the second step, we recompute the light curve and SED by incorporating the four-episode FRB injection while varying the value of $f_{\rm FRBt}$.
      We find for the range of $f_{\rm FRBt}$ for which the modeled light curves, at epochs after the first 1.5 GHz observation, are consistent with the observational data within their uncertainties at all frequencies. 
      This process is then carried out for various values of $\gamma_{\rm b}$, for both the CCSN and USSN cases.
      As we'll see below, the short-timescale flux fluctuations observed over intervals of a few days at the same frequency (Figure \ref{fig:frb_prs_data}) are difficult to reproduce with the four-episode FRB injection model. Therefore in this study, we treat the month-scale flux evolution seen in the 1.5 GHz PRS data as intrinsic variability to be reproduced by the model, and the origin of day-scale variability will be discussed in Section \ref{subsec:prs_flux_variations}.

    \subsection{FRB Emission Properties Constrained by PRS Flux Variations}
    \label{subsec:frb_emission_properties}

     \begin{figure*}[htbp]
       \centering
       \begin{minipage}[b]{0.49\textwidth}
          \centering
          \includegraphics[width=\linewidth]{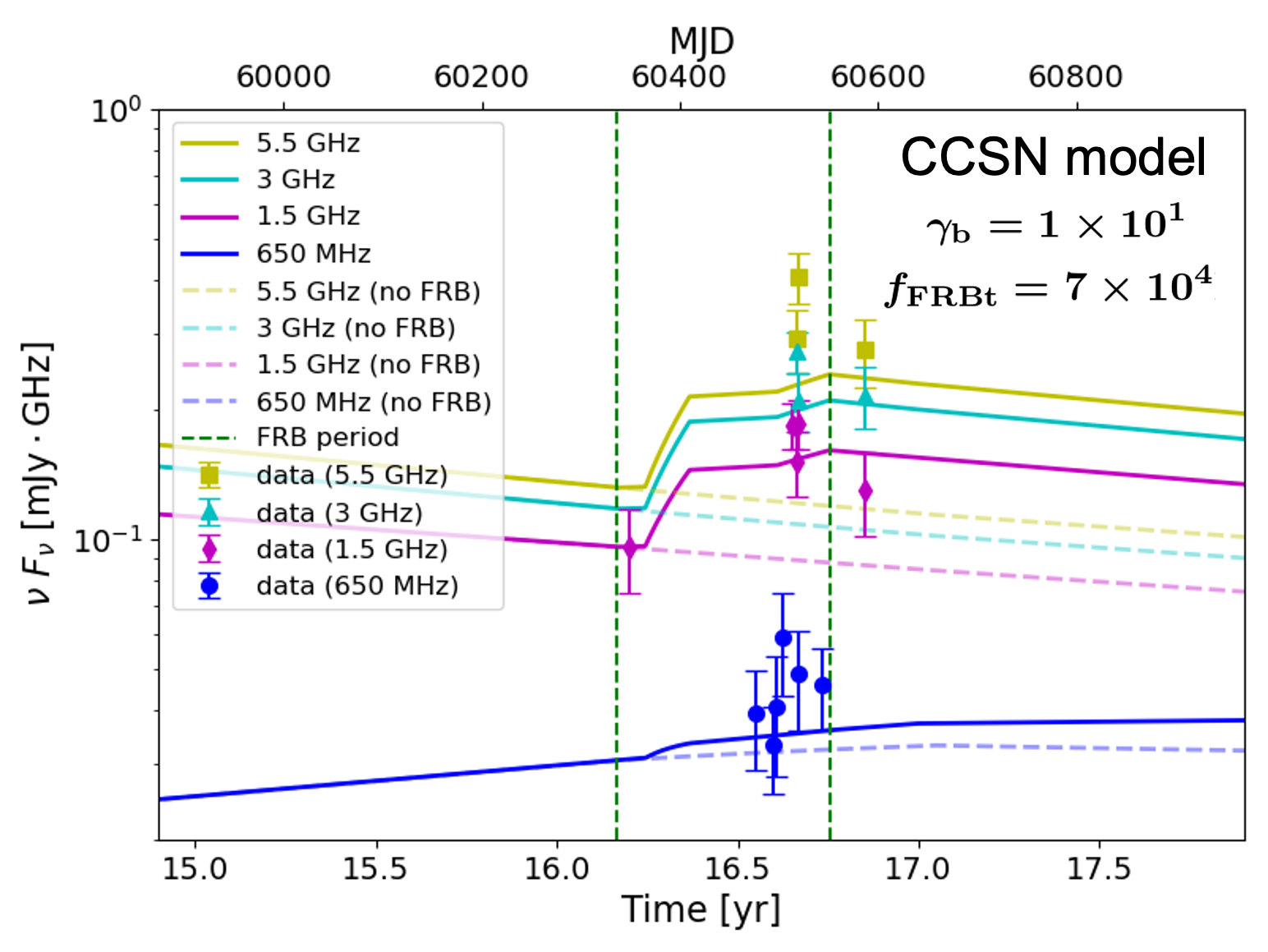}
          \label{figa:ccsn_lightcurve_gamma10}
       \end{minipage}
       \begin{minipage}[b]{0.49\textwidth}
          \centering
          \includegraphics[width=\linewidth]{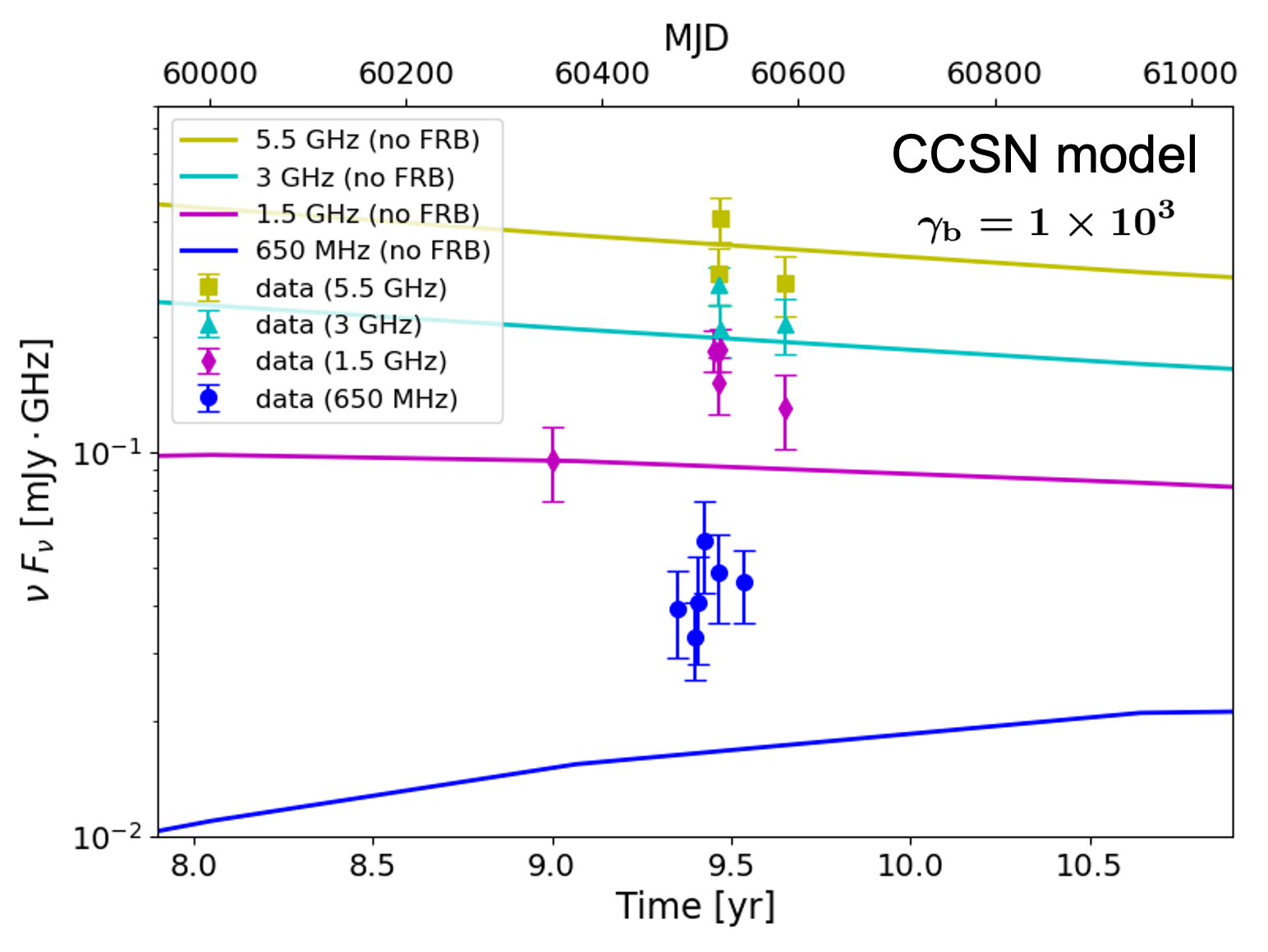}
          \label{figb:ccsn_lightcurve_gamma1000}
       \end{minipage}
       \caption{The model-predicted PRS light curves at 650 MHz, 1.5 GHz, 3 GHz, and 5.5 GHz for the CCSN model, together with the observational data.
       \textit{Left panel}: The PRS light curves with $\gamma_{\rm b}=1 \times 10^1$, $B_{\rm int}=2.0 \times 10^{16}\,{\rm G}$, $t_{\rm age}=16.2\,{\rm yr}$, and $f_{\rm FRBt}=7 \times 10^4$.
       The dashed curves represent the light curves calculated without the four-episode FRB injection, shown for comparison.
       The period of the four-episode FRB injection is indicated by the two vertical green dashed lines.
       \textit{Right panel}: The PRS light curves with $\gamma_{\rm b}=1 \times 10^3$, $B_{\rm int}=1.7 \times 10^{16}\,{\rm G}$, and $t_{\rm age}=9.0\,{\rm yr}$ before introducing the four-episode FRB injection.}
       \label{fig:ccsn_lightcurve_gamma10_1000}
      \end{figure*}

     \begin{figure*}[htbp]
       \centering
       \begin{minipage}[b]{0.49\textwidth}
          \centering
          \includegraphics[width=\linewidth]{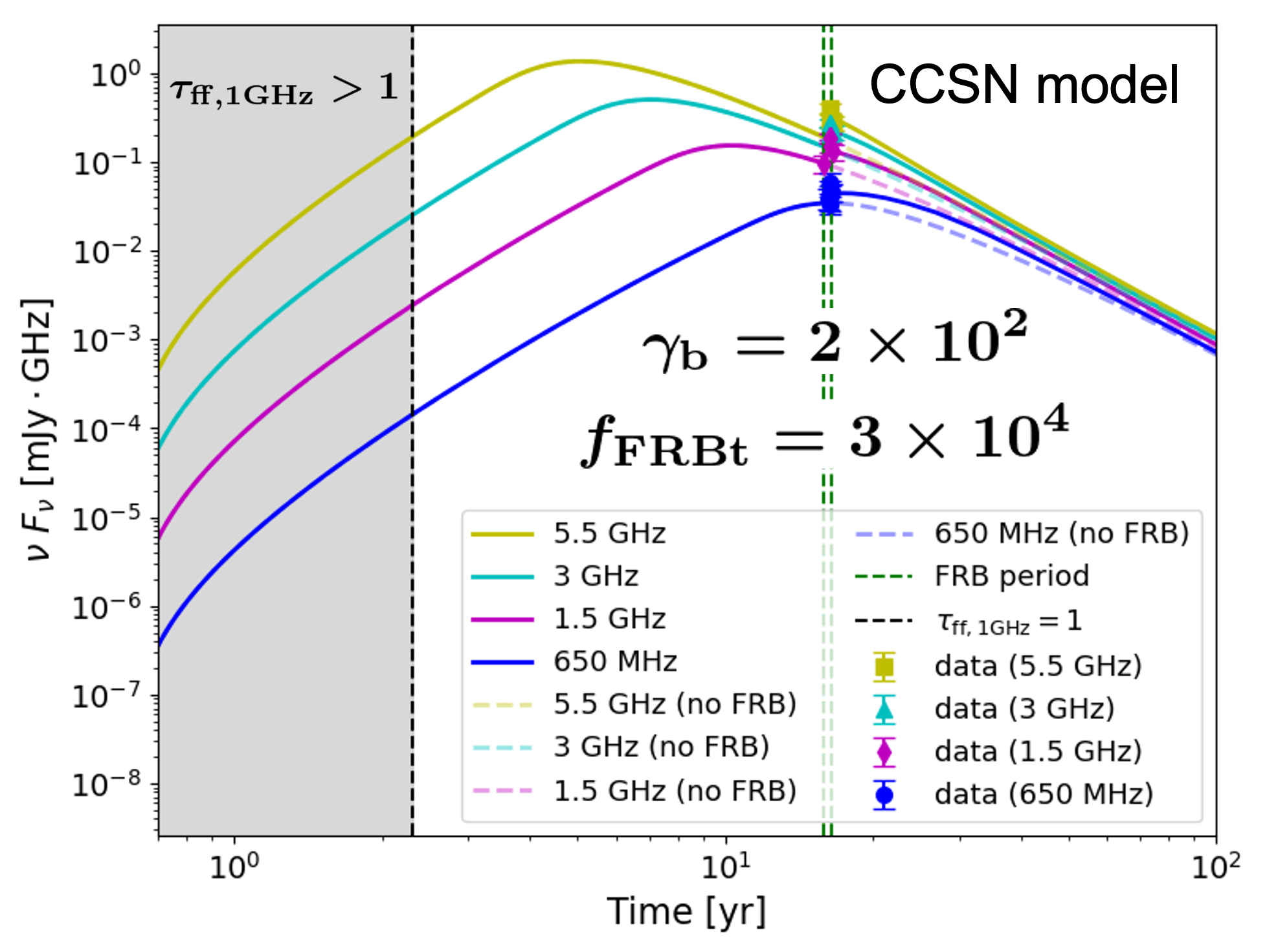}
          \label{figa:ccsn_lightcurve_full}
       \end{minipage}
       \begin{minipage}[b]{0.49\textwidth}
          \centering
          \includegraphics[width=\linewidth]{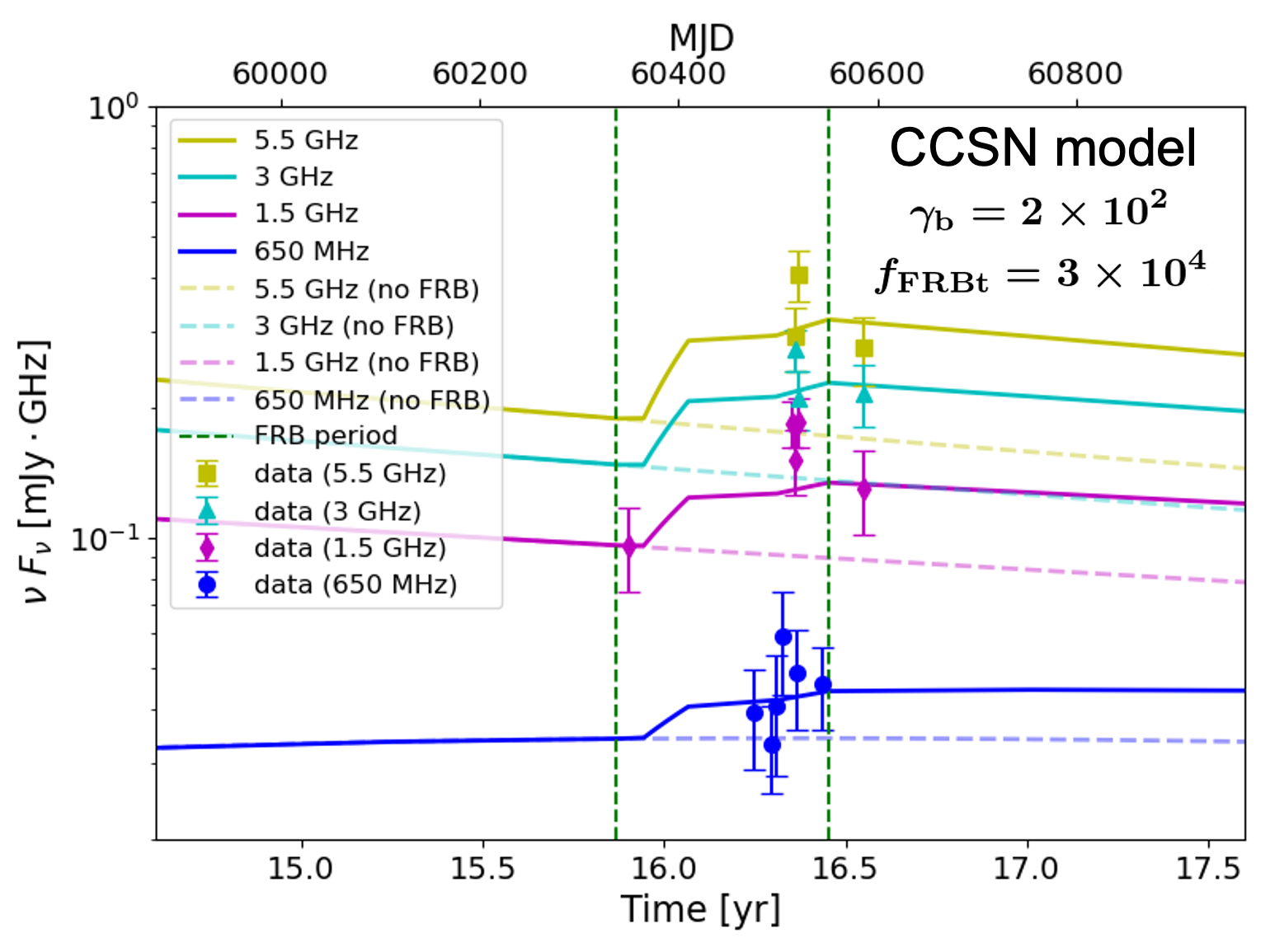}
          \label{figb:ccsn_lightcurve_zoom}
       \end{minipage}
       \caption{The model-predicted PRS light curves using the representative parameters ($\gamma_{\rm b} = 2 \times 10^2$, $B_{\rm int} = 1.3 \times 10^{16} \, {\rm G}$, $t_{\rm age} = 15.9 \, {\rm yr}$, $f_{\rm FRBt}=3 \times 10^4$) for the CCSN model, together with the observational data.
          The dashed curves and the two vertical green dashed lines are the same as those shown in Figure \ref{fig:ccsn_lightcurve_gamma10_1000}.
          \textit{Left panel}: The PRS light curves at 650 MHz, 1.5 GHz, 3 GHz, and 5.5 GHz for the CCSN model. 
          The gray shaded region indicates the epochs where $\tau_{\rm ff, 1 \, GHz} > 1$, during which radio waves are unable to escape from the supernova ejecta.
          \textit{Right panel}: A zoom-in of the epoch of the observed data.}
       \label{fig:ccsn_lightcurve}
      \end{figure*}

      \begin{figure}[htbp]
        \centering
        \includegraphics[width=1.0\linewidth]{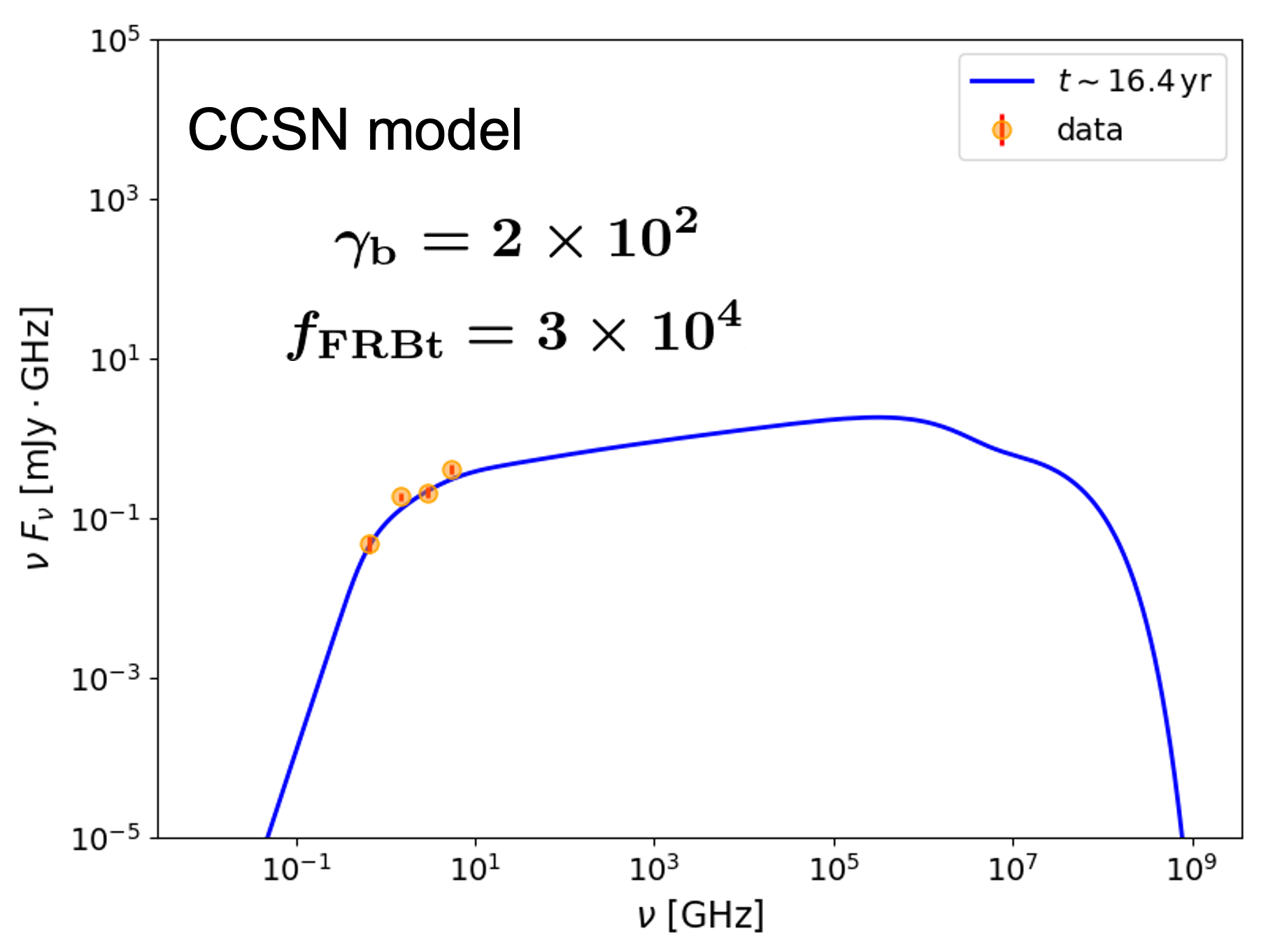}
        \caption{The model-predicted synchrotron SED of the PRS for the CCSN model, together with the observational data (shown by the orange points), using the same set of parameters as in Figure \ref{fig:ccsn_lightcurve}.
         The SED is at the epoch of the observed data (late July 2024), with the corresponding age $t \sim 16.4 \, {\rm yr}$ derived based on $t_{\rm age}$.}
        \label{fig:ccsn_sed}
      \end{figure}

      \begin{figure*}[htbp]
       \centering
       \begin{minipage}[b]{0.49\textwidth}
         \centering
         \includegraphics[width=\linewidth]{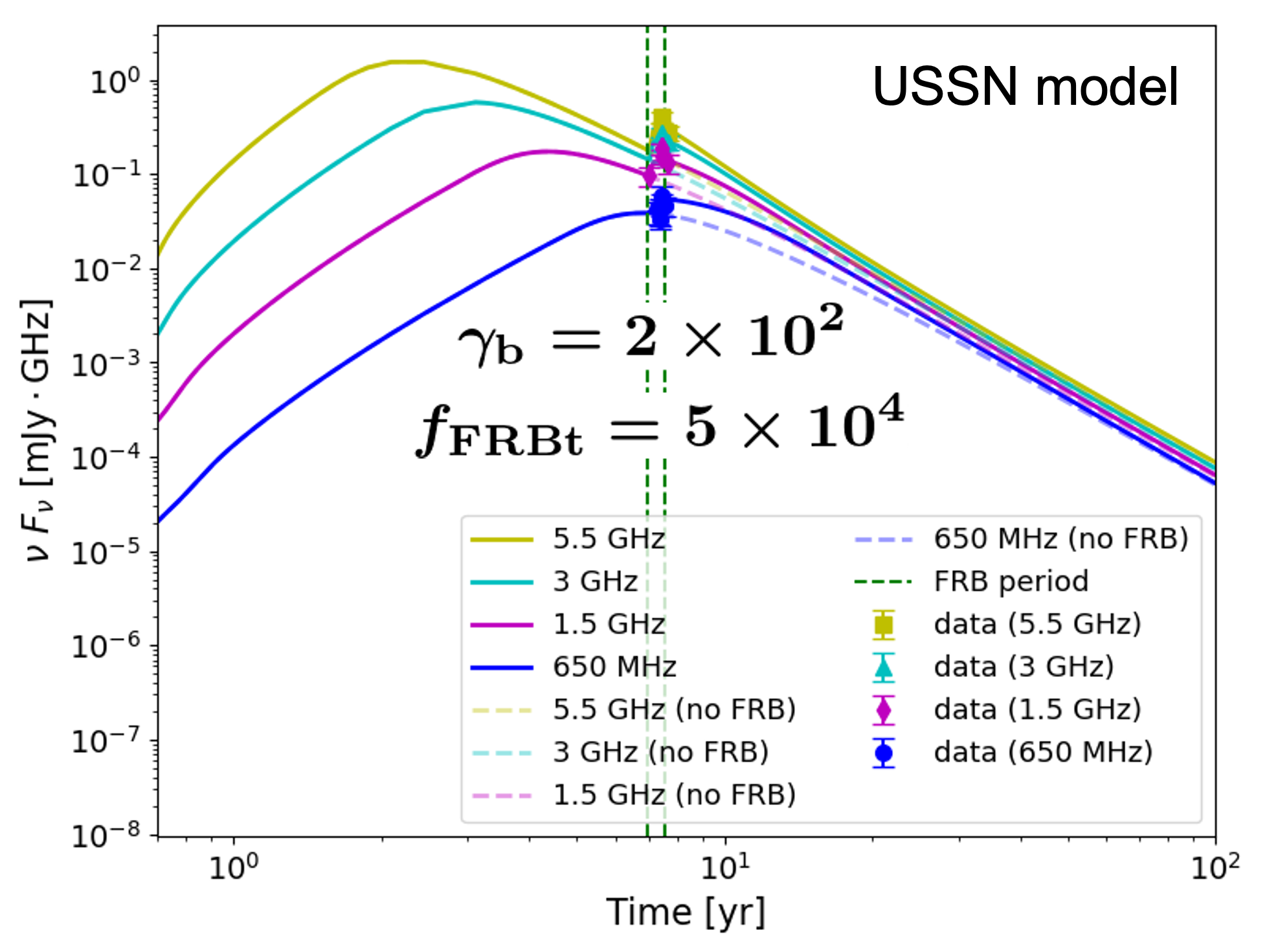}
         \label{figa:ussn_lightcurve_full}
       \end{minipage}
       \begin{minipage}[b]{0.49\textwidth}
         \centering
         \includegraphics[width=\linewidth]{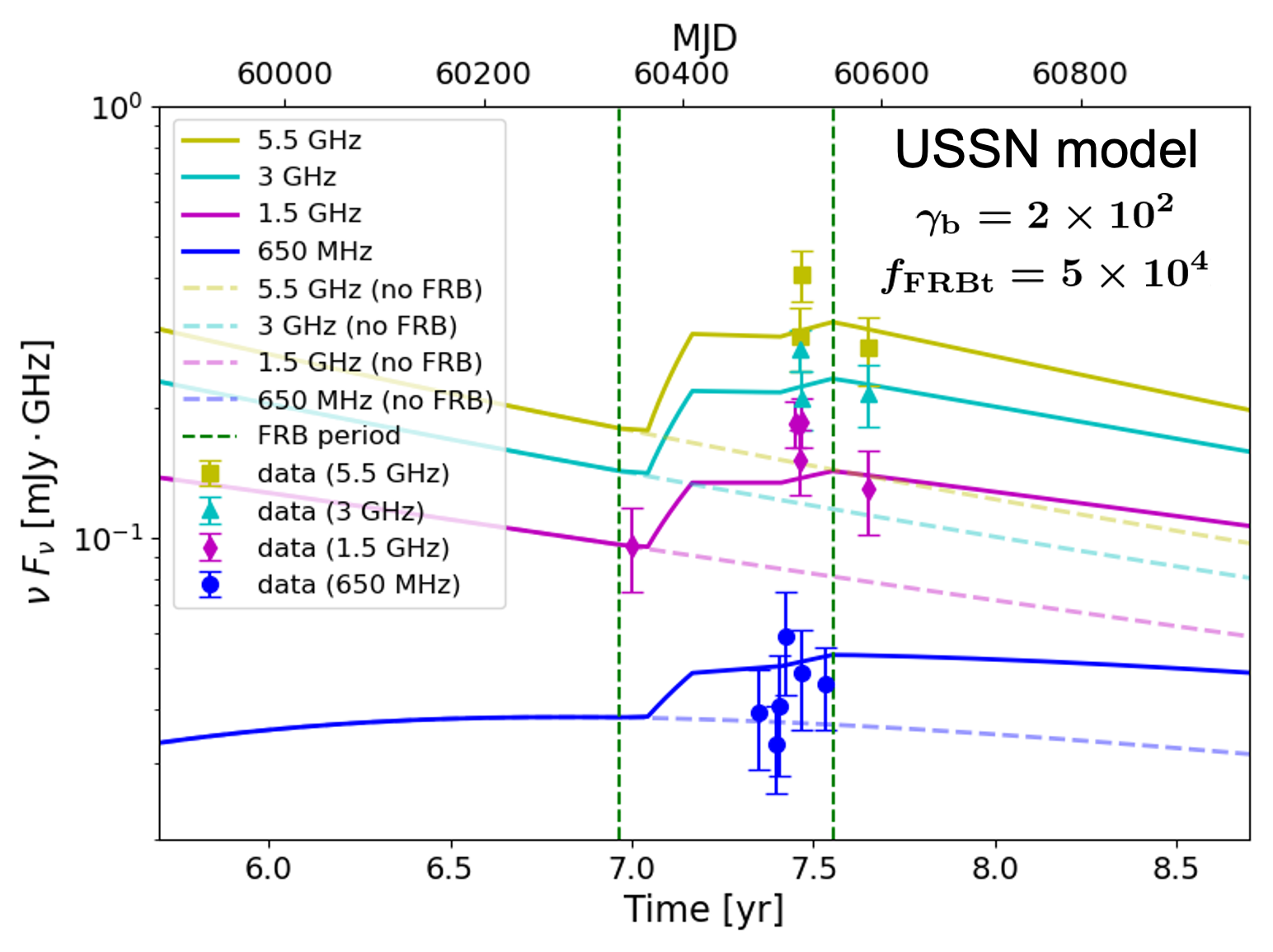}
         \label{figb:ussn_lightcurve_zoom}
       \end{minipage}
       \caption{The model-predicted PRS light curves using representative parameters ($\gamma_{\rm b} = 2 \times 10^2$, $B_{\rm int} = 1.3 \times 10^{16} \, {\rm G}$, $t_{\rm age} = 7.0 \, {\rm yr}$, $f_{\rm FRBt}=5 \times 10^4$) for the USSN model, together with the observational data.
            Other details are the same as in Figure \ref{fig:ccsn_lightcurve}.}
        \label{fig:ussn_lightcurve}
      \end{figure*}

      \begin{figure}[htbp]
        \centering
        \includegraphics[width=1.0\linewidth]{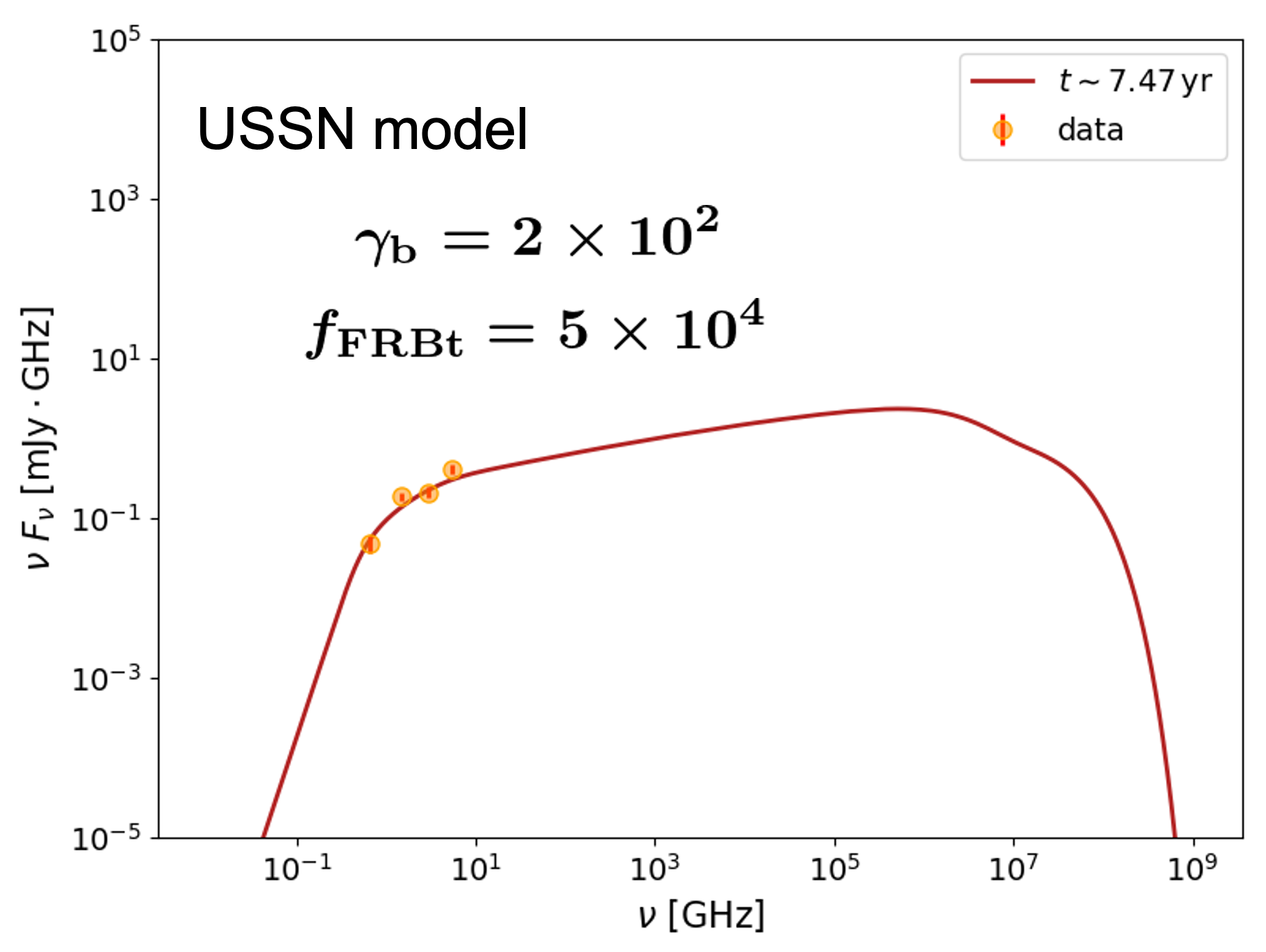}
        \caption{The same as Figure \ref{fig:ccsn_sed}, but for the USSN model using the same set of parameters as in Figure \ref{fig:ussn_lightcurve}.
         The corresponding age is $t \sim 7.47 \, {\rm yr}$ derived based on $t_{\rm age}$.}
        \label{fig:ussn_sed}
      \end{figure}

      \begin{figure*}[htbp]
       \centering
       \begin{minipage}[b]{0.49\textwidth}
         \centering
         \includegraphics[width=\linewidth]{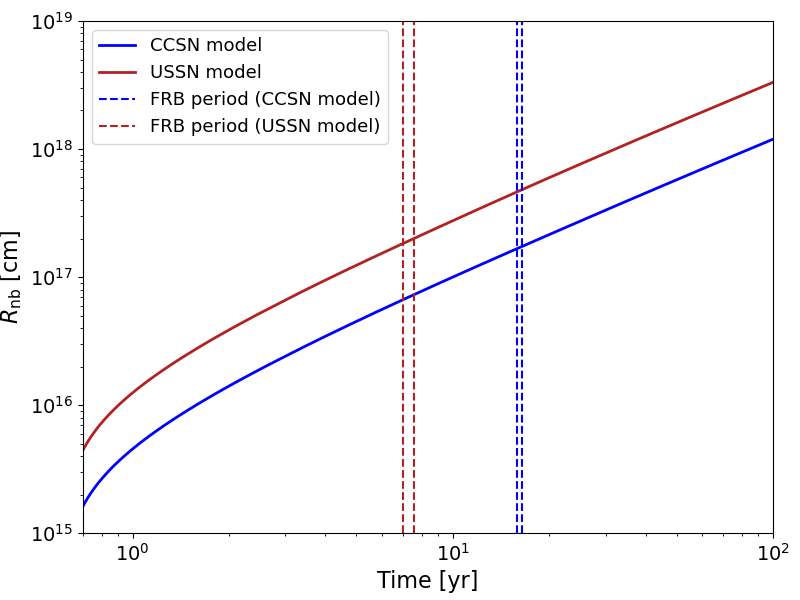}
         \label{figa:shell_radius}
       \end{minipage}
       \begin{minipage}[b]{0.49\textwidth}
         \centering
         \includegraphics[width=\linewidth]{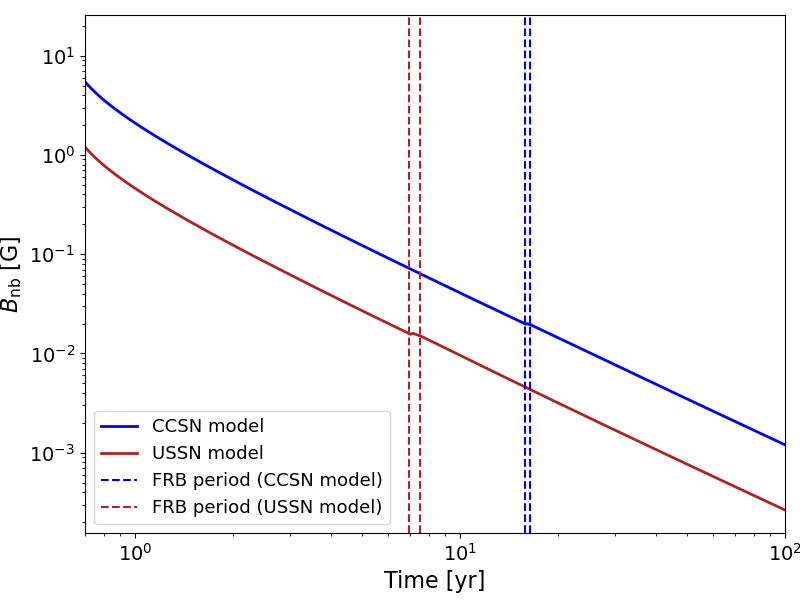}
         \label{figb:b_nb}
       \end{minipage}
       \caption{Time evolution of the MWN radius $R_{\rm nb}$ (left panel) and the average nebular magnetic field $B_{\rm nb}$ (right panel), for the CCSN and USSN models. 
       In both panels, the blue solid curve represents the CCSN model, while the red solid curve represents the USSN model. 
       The blue and red dashed vertical lines indicate the periods of the four-episode FRB injection for the CCSN and USSN models, respectively.
       The model parameters are the same as those used in Figures \ref{fig:ccsn_lightcurve} and \ref{fig:ussn_lightcurve} for the CCSN and USSN models, respectively.}
        \label{fig:nebula_time_evolution}
      \end{figure*}

      Following the fitting procedure described in Section \ref{subsec:fitting_procedure}, we identify parameter sets for both the CCSN and USSN models that are consistent with the observed flux variations of the PRS within the observational uncertainties.

      The left panel of Figure \ref{fig:ccsn_lightcurve_gamma10_1000} presents the PRS light curves in each frequency band (650 MHz, 1.5 GHz, 3 GHz, 5.5 GHz) for the CCSN model with $\gamma_{\rm b}=1 \times 10^1$, for which the best-fit parameters are $B_{\rm int}=2.0 \times 10^{16}\,\mathrm{G}$, $t_{\rm age}=16.2\,\mathrm{yr}$, and $f_{\rm FRBt}=7 \times 10^{4}$.
      Although the modeled PRS flux at 5.5 GHz is slightly lower than the observed value around MJD 60520, the overall light-curve evolution does not show a significant discrepancy from the data, and this parameter set cannot be conclusively ruled out.
      The right panel of Figure \ref{fig:ccsn_lightcurve_gamma10_1000} shows the PRS light curves for the CCSN model with $\gamma_{\rm b}=1 \times 10^{3}$ before introducing the four-episode FRB injection.
      The best-fit parameters are $B_{\rm int}=1.7 \times 10^{16}\,\mathrm{G}$ and $t_{\rm age}=9.0\,\rm{yr}$.
      For $\gamma_{\rm b} \gtrsim 10^{3}$, the synchrotron characteristic frequency corresponding to $\gamma_{\rm b}$ exceeds the observing frequencies, resulting in a harder spectrum and significantly suppressed fluxes at lower frequencies.
      Consequently, the modeled PRS light curves clearly deviate from the observed data, particularly at 650 MHz.

      The same trend is found for the USSN model. Therefore, for both progenitor cases, the allowed range of $\gamma_{\rm b}$ is mildly constrained to values below $10^{3}$.
      For allowed values of $\gamma_{\rm b}$, the observed increase in the PRS flux from MJD 60349 to 60520 is reproduced when $f_{\rm FRBt}$ is of order $10^{4}$--$10^{5}$.
      In both models, $\gamma_{\rm b} \sim 2 \times 10^{2}$ appears to provide the most reasonable agreement with the observed PRS light curves across all frequencies.
      In the following discussion, we adopt $\gamma_{\rm b}=2 \times 10^{2}$ as the fiducial value.

      For the CCSN model, the PRS light curves with $\gamma_{\rm b}=2 \times 10^2$ are shown in Figure \ref{fig:ccsn_lightcurve}, along with the observational data.
      The best-fit parameters are $B_{\rm int}=1.3 \times 10^{16} \, {\rm G}$, $t_{\rm age}=15.9 \, {\rm yr}$, and $f_{\rm FRBt}=3 \times 10^4$.
      The region of $\tau_{\rm ff, 1 \, GHz} > 1$ is indicated in the left panel, and radio waves, including FRBs, can escape through the supernova ejecta after $\sim$ 2 yrs. 
      The model can account for the radio flux increase from MJD 60349 to 60520 in several months by the energy injection associated with the high FRB activity of FRB 20240114A. 
      
      Figure \ref{fig:ccsn_sed} shows the model-predicted synchrotron SED of the PRS for the CCSN model, together with the observational data (shown by the orange points), using the same set of parameters as in Figure \ref{fig:ccsn_lightcurve}.
      The SED corresponds to the epoch at which the observational data were obtained (late July 2024), with the corresponding $t \sim 16.4 \, {\rm yr}$ derived based on $t_{\rm age}$.
      The indication of a SSA break seen in the data, as reported by \citealt{zhang_2025b}, can be reproduced by the synchrotron SED.

      The results for the USSN model with $\gamma_{\rm b} = 2 \times 10^2$ are shown in Figures \ref{fig:ussn_lightcurve} and \ref{fig:ussn_sed}, using the same format as Figures \ref{fig:ccsn_lightcurve} and \ref{fig:ccsn_sed}, respectively.
      The best-fit parameters are $B_{\rm int} = 1.3 \times 10^{16} \, {\rm G}$, $t_{\rm age} = 7.0 \, {\rm yr}$, and $f_{\rm FRBt}=5 \times 10^4$.
      Owing to the smaller ejecta mass compared to the CCSN model, the condition $\tau_{\rm ff, 1 \, GHz} < 1$ is already satisfied at $t \sim t_{\rm inj}$.
      As with the CCSN model, it successfully reproduces the increase in the PRS radio flux on a timescale of several months and the SED.    
      The comparison between the model predictions and the observational data suggests no clear difference between the CCSN and USSN models, with the allowed ranges of $\gamma_{\rm b}$ and $f_{\rm FRBt}$ being of the same order.
      The primary difference between the two models is the age of the PRS corresponding to the observational data.
      This arises from differences in the progenitor ejecta mass ($M_{\rm ej}$) and explosion energy ($E_{\rm sn}$), which shift the timing at which the PRS light curve peaks and begins to decline.

      Since the constraints on $f_{\rm FRBt}$ are obtained, the total energy injected into the MWN during the FAST observing period can be estimated as 
        \begin{equation}    
        E_{\rm FRBt} \approx f_{\rm FRBt} \cdot \sum_{i=1}^{4} L_{{\rm r,obs,iso},i} \Delta t_i . 
        \end{equation}
      In that period, given that our fitting yields $f_{\rm FRBt} \sim 10^4$--$10^5$, the total energy injected into the MWN is estimated to be $E_{\rm FRBt} \sim 10^{48} \, {\rm erg}$.

      Figure \ref{fig:nebula_time_evolution} shows the time evolution of the MWN radius, $R_{\rm nb}$, and the average nebular magnetic field, $B_{\rm nb}$, for the CCSN and USSN models with $\gamma_{\rm b} = 2\times10^{2}$. 
      As shown in Figure \ref{fig:nebula_time_evolution}, both models predict that $R_{\rm nb}$ increases with time to $\sim10^{17}\,\rm{cm}$ at the epoch of the PRS observations, while $B_{\rm nb}$ decreases to $\sim10^{-2}\,\rm{G}$ at the same epoch.
      These values allow us to make predictions for the future flux evolution of the PRS, which are discussed in Section \ref{subsec:prs_flux_variations}.

      Given that $R_{\rm nb}\sim10^{17}\,\rm{cm}$ at the epoch of the PRS observations, energy injected from the central magnetar requires $R_{\rm nb}/c \sim 10^{2}$ days to reach the shell at the speed of light.
      Assuming that the FRB-emitting pair plasma propagates through the MWN with a bulk Lorentz factor of $\gamma_{\rm b} \sim 2\times10^{2}$, the arrival-time delay relative to electromagnetic signals is $\sim (R_{\rm nb}/c)/(2\gamma_{\rm b}^2) \sim 10^{2}\,\rm{s}$.
      This timescale is much shorter than the week-to-month PRS flux-variation timescales considered in this study, supporting our model’s assumption that the enhanced FRB activity affects the PRS instantaneously.

  \section{Discussion}
  \label{sec:discussion}
    The fitting results presented in Section \ref{subsec:frb_emission_properties} suggest that the indications of flux variability in the PRS can be exploited to place constraints on the intrinsic energy of FRBs, as well as on the physical properties of the plasma responsible for their production.
    Based on the allowed ranges of parameters obtained in this study, we discuss the validity of the magnetar model and the interpretation in which the PRS flux variability reflects the activity of FRBs.

    \subsection{Energy Budget of the Magnetar Required to Power FRBs}
    \label{subsec:magnetar_energy_budget}
      It has been argued by \citealt{zhang_2025a} that, for highly active sources such as FRB 20240114A, the total magnetic energy of a typical magnetar may be insufficient to power the observed FRB activity.
      Furthermore, following the detection of $\sim$ 11,000 FRBs with FAST in 2024, subsequent monitoring of FRB 20240114A by the HyperFlash and \'ECLAT programs up to this year  (\citealt{ould-boukattine_2026}) has revealed a cumulative burst energy reaching approximately twice that derived from the FAST sample.
      However, the above argument regarding the magnetar energy budget is based only on fiducial assumptions for the FRB parameters, namely $b_{\rm FRB}=0.1$ and $\eta_{\rm r} =10^{-4}$ \citep{zhang_2025a}, together with an assumed magnetar magnetic field strength of $\sim 10^{15}\,\rm{G}$. Thus, it remains unclear whether a magnetar can realistically power the observed activity of FRB 20240114A.
      It is therefore necessary to quantitatively examine the adequacy of the magnetar energy budget, taking into account the intrinsic energy released in individual FRB bursts derived in this study.

      In the CCSN model, by the onset of the four-episode FRB injection ($t_{\rm age} = 15.9\,\mathrm{yr}$), the cumulative magnetic energy dissipated through the time-decaying energy injection rate, $L_{\rm mag}(t)$, reaches $E_{\rm inj}(t_{\rm age}) \sim 3.0 \times 10^{49}\,\mathrm{erg}$. 
      This corresponds to about $60 \, \%$ of the initial magnetic energy budget, $E_{\rm mag,i} \sim 4.9 \times 10^{49}\,\mathrm{erg}$, for a magnetar with $B_{\rm int} = 1.3 \times 10^{16}\,\mathrm{G}$. 
      During the four-episode injection phase, the total energy injected into the MWN is $\sim 10^{48}\,\mathrm{erg}$, estimated from the allowed range of $f_{\rm FRBt}$, and is thus well within the available energy budget.
      In the USSN model, a similar conclusion is reached. 
      By $t_{\rm age} = 7.0\,\mathrm{yr}$, the injected energy amounts to $E_{\rm inj}(t_{\rm age}) \sim 2.5 \times 10^{49}\,\mathrm{erg}$, corresponding to about $50 \, \%$ of the initial magnetic energy budget for $B_{\rm int} = 1.3 \times 10^{16}\,\mathrm{G}$. 
      The total FRB energy released during the injection phase is comparable to that in the CCSN model, $\sim 10^{48}\,\mathrm{erg}$.

      Overall, in both models, the observed level of FRB activity can be naturally accommodated within the magnetic energy reservoir of the magnetar. 
      Taking into account the remaining energy budget, our results suggest that FRB 20240114A could sustain a similar level of activity for of order $10$ additional episodes comparable to the four-episode FRB injection considered in this study.

      As discussed in Section \ref{subsec:magnetar_energy_injection}, there is substantial degeneracy in the combination of $t_{\rm inj}$ and $\alpha$, which determine the magnetic energy injection $L_{\rm mag}(t)$ [Equation (\ref{magnetic_flare_luminosity})]. 
      As a result, the inferred age of the PRS corresponding to the observational data can vary depending on the adopted parameter set. 
      However, we have confirmed that varying $t_{\rm inj}$ and $\alpha$ by factors of a few primarily changes the inferred system age, while having only a minor impact on the cumulative magnetic energy dissipated up to the epoch of the PRS observations. 
      Therefore, the discussion on the total energy budget is not seriously affected by these uncertain parameters.

      We also note that the magnetic field amplification efficiency, $\epsilon_B$, introduced in Equation (\ref{average_magnetic_field}), is uncertain at the order-of-magnitude level. 
      Since $B_{\rm nb}$ depends on $\epsilon_B$, the characteristic Lorentz factor of the electrons responsible for the PRS emission also varies with $\epsilon_B$. 
      Consequently, the age at which the PRS light curve transitions from a rising to a declining phase depends on the adopted value of $\epsilon_B$. 
      We have confirmed that varying $\epsilon_B$ by an order of magnitude shifts this transition age by several years, resulting in a comparable change in the inferred system age. 
      However, its effect on the energetics at the epoch of the PRS observations is found to be minor, as the cumulative magnetic energy injected by that epoch changes by less than $10 \, \%$ when $\epsilon_B$ is varied by an order of magnitude.

      Another parameter affecting the discussion of the magnetar energy budget is the injection efficiency, $\epsilon_e$ introduced in Equation (\ref{injection_rate}). 
      In this study, we assume that the injected energy is efficiently converted into a population of non-thermal electrons, following previous PRS modeling studies. 
      However, if this conversion efficiency is lower, a larger energy budget from the central engine would be required to supply sufficient high-energy electrons responsible for the PRS emission.
      A more detailed assessment of such microphysical parameters will be essential for placing tighter constraints on the energy budget.

      Recent observations of FRB 20240114A by the HyperFlash monitoring program (\citealt{ould-boukattine_2026}) have detected 178 high-energy radio bursts, each with isotropic energies of $E_{\rm iso} \sim 10^{40\mbox{--}42}\,\mathrm{erg}$. 
      Including these events, the cumulative energy released in FRBs to date is estimated to be approximately twice that inferred from the FAST sample alone.
      In light of the energy budget considerations discussed above, the magnetar scenario is not immediately ruled out at present. 
      However, if bursts with energies comparable to those observed by HyperFlash continue to occur, this will place stringent constraints on the magnetar origin of FRB 20240114A.

      \subsection{On the duty cycle of the hyper-active phase of FRB 20240114A}
      Focusing on the total energy injection rate, $L(t)$, from the magnetar, we find that the steady magnetic energy injection rate, $L_{\rm mag}$, around the epochs of the four-episode FRB injection is of order $\sim 10^{40}\,{\rm erg\,s^{-1}}$. 
      In contrast, the energy injection rate associated with the activity of FRB 20240114A, $L_{\rm FRBt}$, reaches $\sim 10^{42}\,{\rm erg\,s^{-1}}$ during the 2nd episode, corresponding to the most hyper-active phase of FRB 20240114A during the FAST monitoring period.
      An approximate constraint on the duty cycle of hyper-active phases can be derived if the steady injection component, $L_{\rm mag}$, is also powered by magnetic energy release associated with FRB activity.
      In order for the time-averaged energy production during the hyper-active phase not to exceed $L_{\rm mag}$, the duty cycle must be approximately $1 \, \%$ or less.
      Equivalently, the probability of observing FRB 20240114A in such an extreme hyper-active state is expected to be no larger than about $1 \, \%$.

      This means that if observations spanning a period of several months are repeated 100 times, a hyper-active phase, such as the second episode of FRB 20240114A, will be observed only once.
      To date, approximately ten confirmed and candidate FRB-PRS systems (see Section \ref{sec:intro}) have typically been monitored for several years, which correspond to approximately $100$ independent few-month monitoring intervals.
      Assuming that all of these FRB-PRS systems are powered by young magnetars embedded in MWNe (Figure \ref{fig:schematic_picture}), the fact that the hyper-active phase of FRB 20240114A was observed to be the most active one yet is therefore broadly consistent with the duty-cycle estimate discussed above.

    \subsection{Interpretation of the PRS Flux Variations and Predictions for Future Observations}
    \label{subsec:prs_flux_variations}
      The fitting results suggest that the observed flux variations of the PRS on timescales of months can be explained by the four-episode FRB injection model, in which the activity of FRBs directly influences the PRS emission.
      However, as shown in the right panels of Figures \ref{fig:ccsn_lightcurve} and \ref{fig:ussn_lightcurve}, the PRS light curve does not immediately decline even if the FRB activity becomes less active.
      This behavior is due to the fact that the synchrotron cooling timescale of the MWN is much longer than the observation period.
      For both the CCSN and USSN models, the average magnetic field strength in the MWN at time of the PRS observations $(t \sim t_{\rm age})$ is $B_{\rm nb} \sim 10^{-2}\,\rm{G}$, as shown in the right panel of Figure \ref{fig:nebula_time_evolution}. 
      The typical Lorentz factor of electrons responsible for the PRS radio emission at $\bar{\nu} \sim 1\,\mathrm{GHz}$ is $\bar{\gamma} \approx (4\pi m_e c \bar{\nu} / 3 e B_{\rm nb})^{1/2} \sim 10^{2}$, resulting in the synchrotron cooling timescale of $t_{\rm syn} \approx 6\pi m_e c / \sigma_{T} \bar{\gamma} B_{\rm nb}^2 \sim 10^3 \, \rm{yr}$.     
      Under these conditions, adiabatic cooling is expected to dominate in this phase, and electrons cool on the dynamical timescale of the MWN (of order a few decades).
      These cooling timescales act to smooth out the temporal response of the PRS emission, and the PRS light curve does not exhibit an immediate decline on month-scale observational timescales, even after the FRB activity subsides.

      Although part of the observed PRS flux variations may be partially caused by refractive scintillation in the Galactic ISM, the characteristic refractive timescale is estimated to be a few days (\citealt{zhang_2025b}).
      Hence, while the fine week-scale fluctuations observed at 1.5 GHz may plausibly originate from scintillation, the longer-term month-scale evolution of the PRS flux may involve an intrinsic origin.

      \citealt{li_2026} reported that the PRS flux of FRB 20240114A varies on week-to-month timescales in apparent correlation with the FRB activity. 
      However, in examining this correlation, the PRS flux measurements at different frequencies were converted to a common reference frequency of 3 GHz, assuming a spectral index of $-0.34$ (\citealt{bruni_2025}) inferred from a limited observational band in the $1$--$5 \, \rm{GHz}$ range. 
      Therefore, the robustness of the inferred correlation remains subject to further investigation.

      Our PRS model predicts that plasma injection during a hyper-active phase, such as that of FRB 20240114A, leads to a sharp rise in the PRS light curve, followed by a slow decline over a timescale of several tens of years once the FRB activity diminishes.
      If future observations reveal a clear intrinsic decline in the PRS light curve on week-to-month timescales, it would suggest that the synchrotron cooling timescale must be comparable to these observational timescales, and the current model would need to be revised.
      For example, one possible modification is to invoke strong collimation of the FRB outflows, which could locally enhance the magnetic field strength in the MWN and thereby shorten the synchrotron cooling timescale. 
      More generally, such a scenario would require relaxing the assumption of spherical symmetry adopted in this study.

    \subsection{Constraints on FRB Beaming Properties from PRS Flux Variability}
    \label{subsec:beaming_properties}
      From our analysis, we find that the parameter $f_{\rm FRBt}$, defined as $f_{\rm FRBt} \equiv b_{\rm FRB} \cdot f_{\rm on-axis}^{-1} \cdot \eta_{\rm r}^{-1}$, is of order $10^{4}$--$10^{5}$.
      If FRB emission originates from multiple independent emitting regions, as suggested by \citealt{manaswini_2026}, a completely random beam orientation would imply $b_{\rm FRB} \cdot f_{\rm on-axis}^{-1} \sim 1$, corresponding to $\eta_{\rm r} \sim 10^{-5}$--$10^{-4}$.
      Therefore, the results presented in Section \ref{subsec:frb_emission_properties} suggest that individual FRBs are produced by relativistic outflows with characteristic bulk Lorentz factors of $\gamma_{\rm b} \lesssim 10^{3}$ and the radio radiation efficiency of $\sim 0.001$--$0.01 \, \%$.

      Notably, similar values of the outflow bulk Lorentz factors and radio radiation efficiencies have previously been inferred from studies of PRSs associated with non-active FRBs (e.g., \citealt{beloborodov_2017}). 
      The present work independently reaches a consistent conclusion from the flux variability associated with the hyper-active phase of FRB 20240114A.
      This demonstrates that the PRS flux variability can provide valuable constraints on the beaming properties, and ultimately the emission mechanism of FRBs.

  \section{Summary}
  \label{sec:summary}

    In this study, we investigated whether the observed PRS flux variations around FRB 20240114A can be explained by synchrotron emission from plasma injected into the MWN by magnetar-driven FRBs.
    We also examined whether the emission properties of FRBs can be constrained by linking the FRB activity with the PRS flux variability.

    By adopting a model in which plasma is injected into the MWN in association with the observed FRB activity, we find that the month-scale flux variability of the PRS can be reproduced within the observational uncertainties, except for short-term fluctuations on day-to-week timescales.
    Furthermore, fitting the PRS light curve enables us to constrain not only the magnetar and MWN parameters, such as the system age, but also the physical properties as a FRB source, including the bulk Lorentz factor of the FRB-emitting plasma, as well as the total energy injection into MWN associated with FRBs. 
    This suggests that the observed behavior of the PRS provides a valuable means of probing the emission mechanisms of FRBs.

    Our fitting suggests that, if the magnetar progenitor is a CCSN, the age of the magnetar at the time of FRB 20240114A is $\sim 16\, \rm{yr}$, whereas in the USSN case it is $\sim 7\, \rm{yr}$.
    In either scenario, the results indicate that the FRB activity is powered by a young magnetar with an age of a few tens of years.
    However, these age estimates remain subject to uncertainty, as they depend on the magnetic energy injection rate associated with magnetar flares.
    Although not tightly constrained, the bulk Lorentz factor of the FRB-emitting electrons is favored to be below $10^{3}$ in both progenitor cases.
    In both models, $\gamma_{\rm b} \sim 2 \times 10^{2}$ appears to provide the most reasonable agreement with the observed PRS light curves across all frequencies.
    The total energy injected into the MWN during the FAST observing period is estimated to be $\sim 10^{48}\,\mathrm{erg}$.
    Assuming $\gamma_{\rm b}=2.0 \times 10^2$, the internal magnetic field of the magnetar is estimated to be $B_{\rm int} = 1.3 \times 10^{16}\,\mathrm{G}$ for the both progenitor cases, corresponding to a magnetic energy budget of $\sim 5 \times 10^{49}\,\mathrm{erg}$. 
    Therefore, the magnetar energy budget is sufficient to power both the hyper-active FRB emission and the associated PRS flux variability.

    Our PRS light-curve modeling shows that, in systems with ages of a few tens of years, adiabatic cooling of electrons in the MWN dominates over synchrotron cooling.
    As a result, after the PRS is brightened by active FRB injection, it is predicted to gradually decline on a dynamical timescale of several tens of years once the FRB activity subsides.
    Assuming that the steady magnetic-energy injection into the MWN originates from the same magnetic energy reservoir responsible for FRB production, considerations on energetics predict that the hyper-active phase observed in FRB 20240114A, characterized by burst rates of up to $\sim 700 \, \rm hr^{-1}$, has an occurrence probability (or duty cycle) of at most $\sim 10^{-2}$.
    This duty-cycle estimate is broadly consistent with the fact that only one such hyper-active phase has been identified in the current observational sample of FRB-PRS systems.
    Future radio monitoring of FRB 20240114A and other FRB-PRS systems will provide direct tests of these predictions, allowing to assess the validity of the model presented by this work.

  \begin{acknowledgments}
    This work was supported by the JSPS/MEXT KAKENHI Grant Number 23H04899, 24K00668, and 25K00021 (KK) and 18K03692, 26H00822, 26H02063, and 26K22348 (TT).
  \end{acknowledgments}

  \bibliographystyle{aasjournalv7.1}
  \bibliography{references}

\end{document}